\documentclass[acmtog]{acmart}

\usepackage{amsthm}
\usepackage{color,hyphenat,balance}
\usepackage[fleqn,tbtags]{mathtools}
\usepackage{autonum}
\usepackage{wrapfig}
\usepackage{contour}
\usepackage{microtype}
\usepackage{multirow}
\usepackage{soul}
\usepackage{algorithm}
\usepackage{xspace}
\usepackage{algpseudocode}
\usepackage{diagbox}
\usepackage{enumitem}

\usepackage{mathtools}
\usepackage{lineno}
\acmSubmissionID{718} 

\copyrightyear{2026}
\acmYear{2026}
\setcopyright{cc}
\setcctype{by-nc-nd}
\acmConference[SIGGRAPH Conference Papers '26]{Special Interest Group on Computer Graphics and Interactive Techniques Conference Conference Papers}{July 19--23, 2026}{Los Angeles, CA, USA}
\acmBooktitle{Special Interest Group on Computer Graphics and Interactive Techniques Conference Conference Papers (SIGGRAPH Conference Papers '26), July 19--23, 2026, Los Angeles, CA, USA}
\acmDOI{10.1145/3799902.3811125}
\acmISBN{979-8-4007-2554-8/2026/07}

\begin{document}

%\title{LoBoFit: \\ Detail-Preserving Garment Refitting via Local Bone Projections}
\title{LoBoFit: Flexible Garment Refitting via Local Bone Mapping Blending
}

\author{Meng Zhang}
\authornote{M. Zhang is the corresponding author.}
\email{lynnzephyr@gmail.com}
\orcid{0000-0003-2384-0697}
\affiliation{%
\institution{School of Computer Science and Engineering, Nanjing University of Science and Technology}\country{China}}

\author{Yu Xin}
\orcid{00009-0006-7433-1998}
\affiliation{%
\institution{University of Science and Technology of China}\country{China}}

\author{Feiya Guo}
\orcid{0009-0006-4348-0771}
\affiliation{%
\institution{Nanjing University of Science and Technology}\country{China}}

\author{Kaizhang Kang}
\orcid{0009-0004-4381-0510}
\affiliation{%
\institution{King Abdullah University of Science and Technology }\country{Saudi Arabia}}

\author{Mengyu Chu}
\email{mchu@pku.edu.cn}
\orcid{0000-0002-7358-433X}
\affiliation{
\institution{State Key Laboratory of General Artificial Intelligence, Peking University}
\country{China}}

\author{Ruizhen Hu}
\orcid{0000-0002-6798-0336}
\affiliation{%
\institution{Shenzhen University}\country{China}}

% \author{Niloy J. Mitra}
% \email{n.mitra@cs.ucl.ac.uk}
% \orcid{0000-0002-2597-0914}
% \affiliation{ \institution{University College London and Adobe Research} \country{United Kingdom}} 

\renewcommand\shortauthors{Zhang M. et al}

%\renewcommand{\shortauthors}{Zhang, et al.}

\begin{abstract}

    Garment refitting, the task of adapting a garment from a source to a target avatar, must preserve the original design features and fine-scale wrinkles, a challenge exacerbated by significant shape variations and {varying poses without registration to a shared canonical pose}. Existing methods struggle to balance robustness, efficiency, and fidelity of detail: physics-based simulation is costly, data-driven approaches lack generalizability, and geometry optimization in the full vertex space is often ill-conditioned 
    {and prone to local minima with unsatisfactory quality.}
    %, leading to unstable convergence and loss of detail. 
    We identify that a fundamental limitation lies in the representation: deforming garments directly in global coordinates couples vertices non-locally, creating a complex and poorly-structured optimization landscape.
    Therefore, we introduce \emph{LoBoFit}, a robust refitting method built upon a novel \emph{Local Bone Mapping Blending (LoBoMap Blending)} representation. Instead of manipulating global vertex positions, LoBoMap Blending expresses garment geometry as a linear blend of its mappings into local bone coordinate frames.
    This representation is highly expressive and flexible: local bone mappings yield a pose-robust initialization and a well-conditioned parameterization, while blending weights smooth the optimization landscape and broaden the space of plausible solutions for stable convergence with fine-scale detail preservation.
    % \Chu{This representation is highly expressive and flexible: the local bone projections provide a pose-robust initialization and establish a well-conditioned parameter space for optimization, while the blending weights offer the high expressiveness required to capture fine-scale geometric details.}
    % This representation provides a pose-robust initialization and, more importantly, a well-conditioned, bone-local parameter space for optimization. 
    % Chu: I want to highlight that "We get stability from the local frames AND we get details from the weights.
    The subsequent refinement efficiently resolves collisions and preserves details by optimizing localized residuals, effectively decomposing the complex global deformation into manageable subproblems.
    Our experiments demonstrate that LoBoFit reliably refits high-resolution, single- and multi-layer garments across avatars with large shape and topological differences, while faithfully preserving intricate wrinkles and the intended fit style, outperforming state-of-the-art methods in robustness and output quality.
    %\footnote{\Zhang{The term \emph{projection} in the title refers to the geometric change of frame; \emph{mapping} is adopted throughout the text for terminological clarity.}}

\end{abstract}




\begin{teaserfigure}
\centering
\includegraphics[width=\textwidth]
{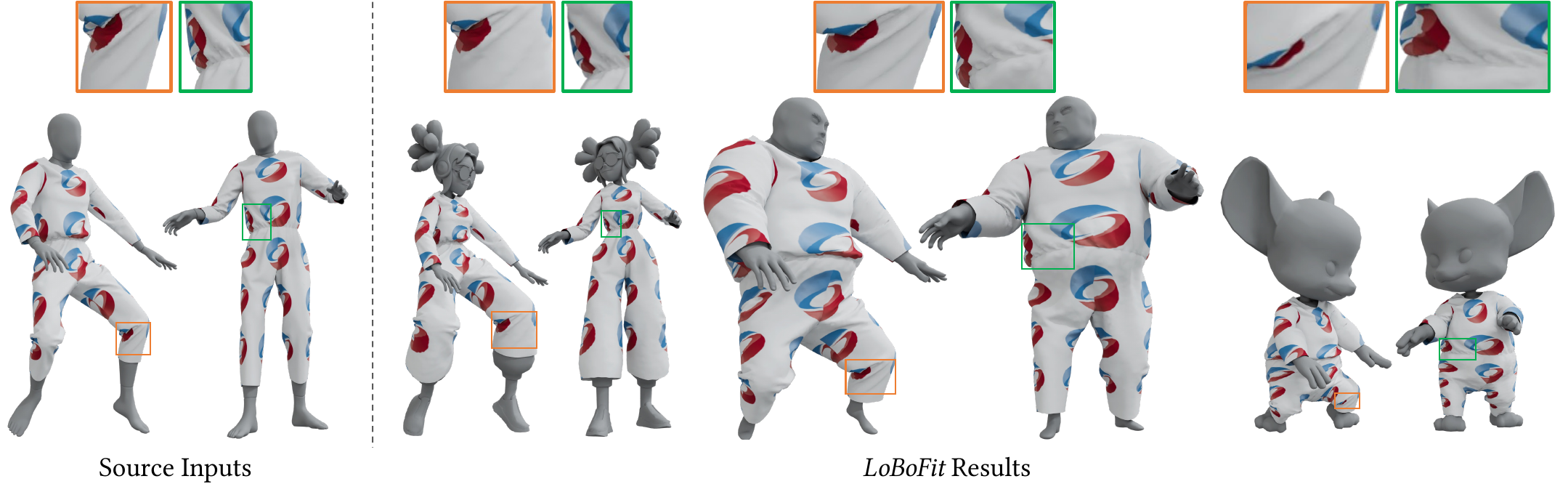}
\vskip -0.15in
\caption{
\emph{LoBoFit} robustly refits a garment designed on a source avatar (any pose) to diverse target builds—from petite to larger—while preserving design features and fine-scale wrinkles.
}  
\label{fig:teaser}
\end{teaserfigure}

\maketitle

\section{Introduction}

Garment refitting—the task of adapting a garment designed for a source avatar to a target body in a shared, arbitrary pose—is {important for applications such as} virtual try-on, game character authoring, and digital film production. While extensive libraries of high-quality garments are created on a limited set of reference characters, practical applications demand efficient adaptation of these assets to avatars with diverse shapes, proportions, and mesh topologies. Crucially, a successful refit must not only be intersection-free but also faithfully preserve the original design intent{, such as the specific tightness of a waistline or the loose silhouette of a bodice, along with the} fine-scale wrinkle details captured in the source pose. This becomes particularly challenging with high-resolution or multi-layer garments, non-canonical poses, and significant topological variations between avatars.

Existing approaches struggle to balance robustness, generality, and detail preservation. Physics-based methods~\cite{bartle2016physics,wang2018rule,chen2025dress} are computationally expensive and often tied to specific parametric body models. Data-driven techniques~\cite{guan2012drape,santesteban2022ulnef,li2023isp,corona2021smplicit,dong2023ag3d} enable fast inference but are limited by their training data, hindering generalization to arbitrary avatars. Geometry-based optimization methods~\cite{Sumner2004Deformation,brouet2012design} offer explicit control but frequently depend on dense correspondences or operate in a poorly-conditioned, full vertex-coordinate space, where optimizing for a global fit leads to correlated, unstable vertex motions that can distort fine details. Recent work like IFGR~\cite{huang2025intersection} reduces correspondence needs via skeletal embedding but relies on hard bone assignments that {are} prone to yield semantically incorrect attachments under non-canonical poses, making the subsequent optimization highly sensitive to initialization.

We observe that a robust refitting pipeline hinges on two key elements: a \emph{pose-invariant initialization} that maintains correct semantic attachment to the target body, and a \emph{well-conditioned parameterization} that decomposes global garment deformation into localized subproblems, 
ensuring stable convergence while remaining flexible and sufficiently expressive to capture fine-scale deformations across diverse shapes, poses, and mesh topologies.
%\Chu{ ensuring stability while offering enhanced flexibility and expressiveness.}
To this end, we introduce \emph{Local Bone Mapping Blending (LoBoMap Blending)}, a novel geometric representation that fundamentally shifts the deformation space from global vertex coordinates to a blend of bone-local mappings.

The core idea is to express a garment vertex not by its world position, but by its coordinates within the local frames of nearby bones. This representation provides a powerful equivariance property: keeping these local coordinates fixed while updating the bone frames 
%(e.g., \Chu{adjusting bone lengths or rotations }for a new body shape) 
(e.g., accounting for changes in bone lengths and joint configurations induced by a different body shape) produces a coherent initial fit on the target avatar. More importantly, it yields a compact optimization space where adjustments are naturally localized to regions influenced by each bone, leading to a more stable and efficient convergence compared to optimizing global vertex offsets.

Building on this representation, we present \emph{LoBoFit}, a complete refitting algorithm. LoBoFit first generates a semantically consistent initial fit via LoBoMap Blending. It then refines the garment through an optimization process that resolves collisions, preserves details via differential coordinates, and maintains a user-specified fit style, all while leveraging a coarse-to-fine strategy for high-resolution garments. As illustrated in Figure \ref{fig:teaser}, our method robustly handles significant shape variations while preserving intricate details.

Our key contributions are:
\begin{itemize}
    \item We introduce \emph{LoBoMap Blending}, a bone-local representation that provides pose-robust initialization and a well-conditioned space for deformation optimization.
    \item We present \emph{LoBoFit}, an optimization-based refitting algorithm that leverages this representation for detail-preserving, collision-aware garment transfer across diverse avatars and poses.
    % \item We present \emph{LoBoFit}, an optimization-based refitting algorithm that leverages this representation for detail-preserving, collision-free garment transfer across diverse avatars 
    % {sharing a wide range of poses.} 
    % and poses.
    \item We demonstrate superior robustness and quality compared to state-of-the-art methods, particularly for non-canonical poses and high-resolution garments with fine wrinkles.
\end{itemize}

\begin{figure*}[t]
    \centering
    \includegraphics[width=\textwidth]{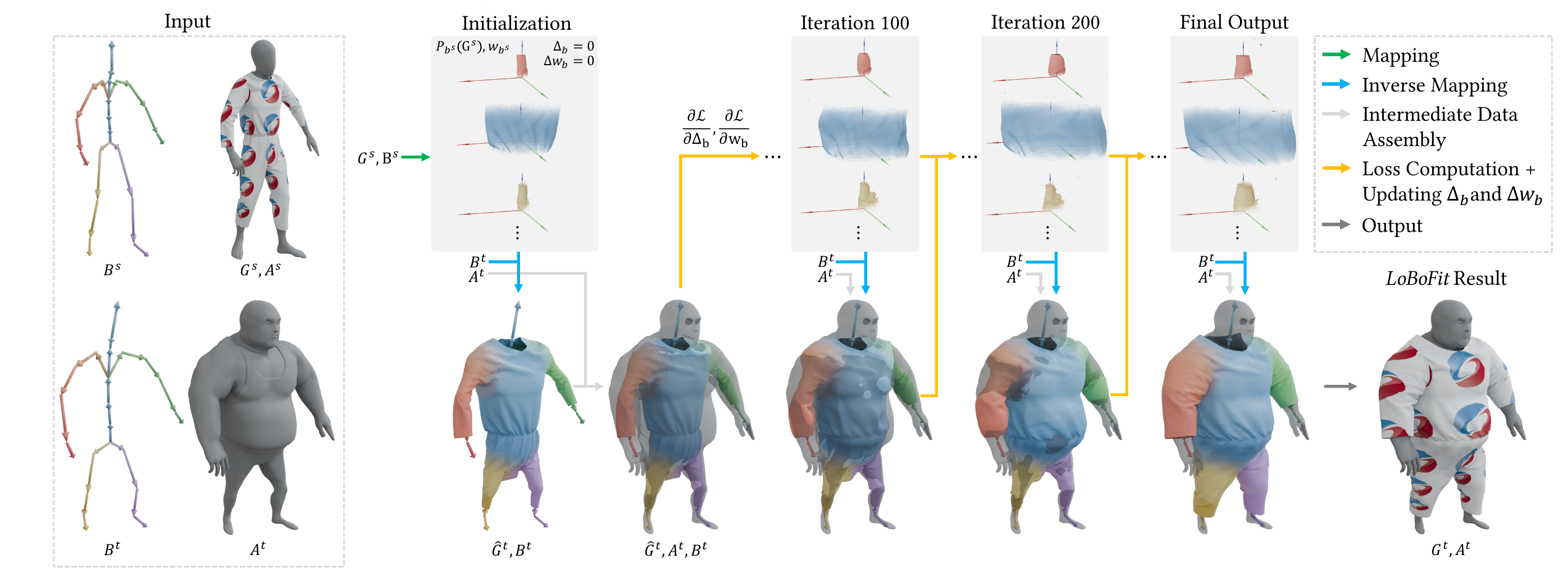}
    \vskip -0.1in
    \caption{ \textbf{\emph{LoBoFit} overview:}
    Given source garment $G^s$ on avatar $A^s$ with bones $B^s$ and target avatar $A^t$ with bones $B^t$, \emph{LoBoFit} initializes $\hat G^t$ by reusing source bone-local coordinates $P_{b^s}(G^s)$ and decoding them with {inverse mappings} on the corresponding target bones. It then optimizes per-bone local-coordinate residuals $\Delta_b$ and weight residuals $\Delta w_b$ by minimizing $\mathcal{L}$, yielding the final refitted garment $G^t$ that preserves design features and fine-scale wrinkles.
%Given a source garment $G^s$ on avatar $A^s$ with rigging bones $B^s$ and a target avatar $A^t$ with rigging bones $B^t$, \emph{LoBoFit} first initializes the refitted garment $\hat G^t$ by reusing the source bone-local projections $P_{b^s}(G^s)$ for each bone $b^s \in B^s$ while decoding them with inverse projections defined on the corresponding target bones $B^t$ .% $B^t$.
%We then iteratively optimize local bone projection residuals $\Delta_b$ and the blending weight residuals $\Delta w_b$ by minimizing the objective $\mathcal{L}$ on the updated target garment mesh. Finally, we obtain the refitted garment $G^t$ on $A^t$, while preserving garment shape, fit style, and fine-scale wrinkle details.
}
    \label{fig:overview}
\end{figure*}

\section{Related Work}

\paragraph{Physics-Driven Refitting Methods} 
Physics-based approaches leverage simulation for realistic garment fitting~\cite{bartle2016physics, guo2021inverse, li2024diffavatar, zheng2024physavatar, grigorev2023hood,grigorev2024contourcraft}.
These methods typically optimize sewing patterns for the rest shape, letting the final fit emerge from physical interactions with the target body and pose.
%These methods typically optimize garment sewing patterns that define the rest shape, while the final fit emerges from physical interactions between the patterns, the target body, and the pose.
\citet{wang2018rule} formulates refitting as a nonlinear optimization problem, but is limited to bodies with small shape variations. Although \citet{eggler2024digital} extend this paradigm to varying body shapes and poses, their method relies on specific parametric models (i.e., SMPL~\cite{SMPL:2015}), which restricts generalization to arbitrary characters. Dress Anyone~\cite{chen2025dress} combines geometric deformation with differentiable cloth simulation, but strictly requires consistent mesh connectivity between source and target bodies. 
Overall, physics-driven pipelines are limited by pattern structure, body parameterization, and simulation cost. We refit garments directly in geometry for efficient transfer across diverse virtual characters, achieving realistic fine-scale details, and leave explicit physical modeling to future work.
%Overall, physics-driven pipelines are constrained by pattern structure, body parameterization, and simulation cost. We sidestep these challenges by operating directly on garment geometry, focusing on efficient geometric refitting for diverse virtual characters while leaving explicit physical modeling to future work.

\paragraph{Data-Driven Garment Models}
Data-driven approaches approximate simulation outcomes conditioned on body shape and pose, yet their generalizability remains fundamentally constrained by training configurations~\cite{kim2008drivenshape, xu2014sensitivity,santesteban2019learning, wang2019learning,santesteban2022snug}. 
Early mesh-based methods like DRAPE~\cite{guan2012drape} and TailorNet~\cite{patel2020tailornet} couple predictions tightly to fixed garment topologies and parametric bodies. %, limiting applicability to arbitrary avatars.
Subsequent work employs image-space~\cite{gu2002geometry, zhang2022motion}, implicit representations~\cite{lahner2018deepwrinkles,santesteban2022ulnef,li2023isp,zhu2022registering}, and generative pipelines~\cite{li2025garmagenet,he2024dresscode}.
While these representations relax mesh-topology constraints, they often sacrifice geometric consistency, controllability, and out-of-domain generalization. In contrast, we use a compact bone-local mapping that preserves explicit geometry and generalizes robustly without a learned prior or a fixed body template.
%While these representations relax mesh-topology constraints for garments, they often trade off geometric consistency, controllability, and generalizability beyond their training domain. Our method instead builds on a compact bone-local projection that retains explicit geometry and ensures robust generalization without requiring a learned prior or a fixed body model.

\paragraph{Geometry-Based Garment Retargeting}
Geometry-based methods retarget garments directly in 3D space. 
Classic deformation transfer techniques~\cite{Sumner2004Deformation, sorkine2004laplacian} rely on explicit correspondences, which are difficult to obtain robustly for diverse garments.
{In particular, the deformation transfer framework of Sumner et al.~\cite{Sumner2004Deformation} is conceptually related, but applying it to garment refitting is non-trivial, as it requires an additional source-to-target body deformation to drive the garment transfer. Some prior garment transfer methods, such as Brouet et al.~\cite{brouet2012design} and Goes et al.~\cite{Goes2020garment} also formulate refitting in geometry space, but rely on dense source--target avatar correspondence or body mapping. Similarly, several more recent works~\cite{pons2017clothcap,lin2024layga, ma2020learning, guo2025progressive} exploit dense avatar correspondences or structured assembly and transfer geometric displacements directly for garment adaptation. These methods are therefore designed under stronger correspondence assumptions than ours.}
%Some prior works~\cite{pons2017clothcap,lin2024layga, ma2020learning} exploit dense avatar correspondences and transfer geometric displacements directly as garment refitting.
%\citet{brouet2012design} formulate garment transfer as a global quadratic optimization, but still require dense correspondence across avatars.
Among recent advances, Intersection-Free Garment Retargeting (IFGR)~\cite{huang2025intersection} is the closest prior work, as it also leverages sparse skeletal correspondences between the source and target avatars. 
However, IFGR uses hard nearest-neighbor bone assignments to shrink the avatar and garment toward the skeleton and reconstructs the garment via IPC-based inflation~\cite{li2020incremental}, which typically requires a clean, intersection-free mesh initialization and can fail under common mesh imperfections such as self-intersections. This dense coupling can further restricts deformation flexibility and increases sensitivity to initialization, especially in non-canonical poses. 
In contrast, we introduce \emph{LoBoFit}, built on a novel garment representation, \emph{LoBoMap Blending}, which softly blends per-bone inverse mappings and optimizes deformation in bone-local subspaces, yielding a pose-robust initialization and a compact, better-conditioned optimization space for stable and efficient convergence. A more detailed comparison is provided in Section~\ref{subsec: compare}.

\section{Overview}
Given a garment $G^s$ originally designed for a source avatar $A^s$ with rigging bones $B^s$,
and a target avatar $A^t$ that shares a similar pose and skeletal topology with rigging bones $B^t$, \emph{LoBoFit} generates a refitted garment $G^t$ that conforms to $A^t$. 
{Unlike prior transfer-based methods~\cite{brouet2012design,Sumner2004Deformation, Goes2020garment}, our method does not require explicit source--target body correspondence.}
We target an efficient end-to-end pipeline for garment retargeting {under varying poses, when the source and target share the same pose, without registration to a canonical pose}.
Figure~\ref{fig:overview} provides a high-level overview of our approach, which consists of two main stages built upon our core representation.

\textbf{Core representation.}
The foundation of our method is \emph{Local Bone Mapping Blending (LoBoMap Blending)}, detailed in Section~\ref{sec: LoBoProj_Blending}. We define a local coordinate frame for each bone in the avatar's skeleton. A mapping function $P_b$ maps any garment vertex into the local coordinates of bone $b$; 
%{here, \emph{projection} is used in a geometric sense to denote this change of frame, rather than a strict linear-algebraic projection}. 
The garment is then reconstructed by linearly blending the inverse {maps} $P_b^{-1}$ of these local coordinates across all bones, weighted by skinning weights $\{w_b\}$ shown in different colors for different bones. This formulation decouples the garment's intrinsic shape (defined by local coordinates) from the global skeleton configuration. 
We illustrate \emph{LoBoMap Blending} in Figure \ref{fig:representation}.

\textbf{Iterative Optimization.}
Building on \emph{LoBoMap Blending}, we present \emph{LoBoFit}, a garment retargeting method detailed in Section~\ref{sec: Garment Refitting}.
We first transfer the source garment $G^s$ to the target avatar $A^t$ by reusing its pre-computed source local coordinates $P_{b^s}(G^s)$ for each bone $b^s \in B^s$. These coordinates are decoded using the inverse {map} $P_{b^t}^{-1}$ (blue arrow) defined on the corresponding target bones $b^t \in B^t$ with the skinning weights $\{w_b\}$. This leverages the equivariance of LoBoMap Blending to instantly produce a high-quality initial garment $\hat{G}^t$ that is semantically attached to the target skeleton.
The initial fit $\hat{G}^t$ is then refined by optimizing two sets of residuals within our representation: \emph{bone-local coordinate residuals} $\Delta_b$ and \emph{blending weight residuals} $\Delta w_b$. 
This optimization (yellow arrow) minimizes a composite loss function $\mathcal{L}$  designed to eliminate garment-body penetrations, preserve the source garment's shape and fine-scale wrinkles, and ensure stable convergence via regularization terms.

\section{LoBoMap Blending} \label{sec: LoBoProj_Blending}
%We introduce \emph{LoBoProj Blending}, which represents garment geometry as a linear blend of per-bone projections.
We introduce \emph{Local Bone Mapping Blending (LoBoMap Blending)}, a geometric representation that encodes garment deformation through bone-local coordinate frames. Unlike traditional methods that manipulate global vertex positions, our representation provides a pose-invariant initialization and a well-conditioned optimization space by decomposing garment geometry into localized bone mappings.

\begin{figure}[t]
  \centering
  %\fbox{\rule{0pt}{4cm}\rule{0.98\columnwidth}{0pt}} % height=4cm, width=0.8\linewidth
  \includegraphics[width=\columnwidth]{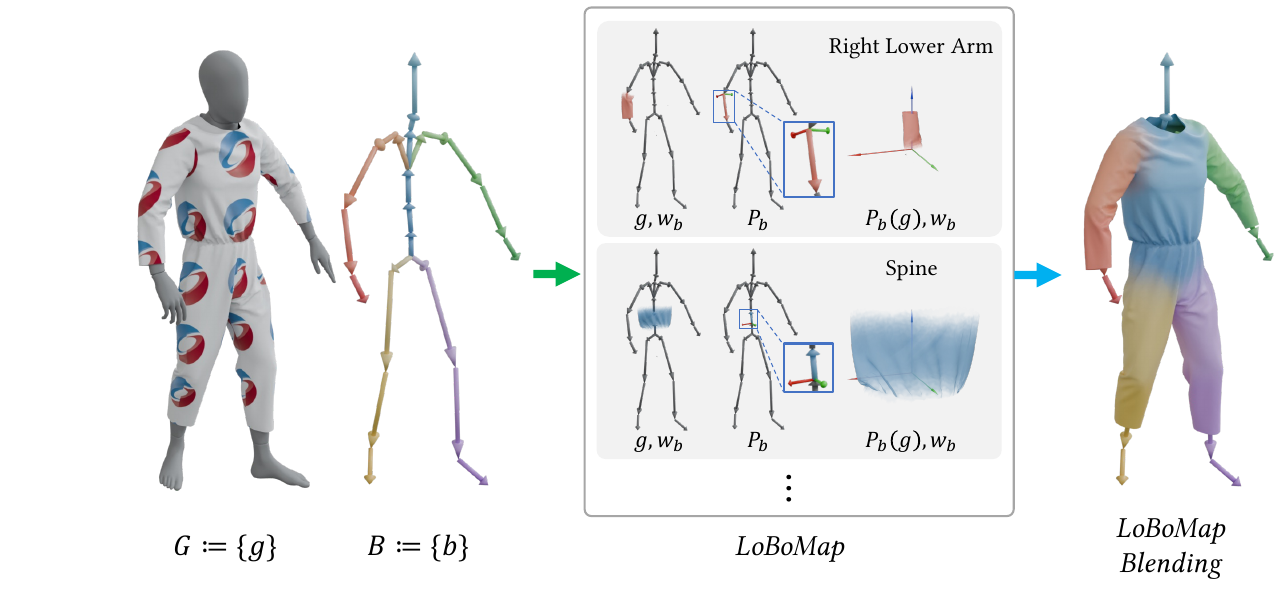}
  \vskip -0.1in
  \caption{\textbf{\emph{LoBoMap Blending}}.
    We represent garment geometry $G$ as a linear blend of per-bone mappings: $P_b$ maps vertices to bone-local frames ($b\in B$), and the garment is reconstructed by blending the inverse {maps} $P_b^{-1}$ using skinning weights ${w_b}$. (RGB encodes the associated bone, and the alpha channel encodes the blending weight ${w_b}$.) 
    %(colors denote bones {and ${w_b}$ is visualized as alpha}).
  % We represent garment geometry $G$ as a linear blend of per-bone projections: $P_b$ maps vertices to bone-local frames ($b\in B$), and the garment is reconstructed by blending the inverse projections $P_b^{-1}$ using skinning weights ${w_b}$ (colors denote bones).
  %We express garment geometry $G$ as a linear blend of per-bone projections. A projection function $P_b$ maps each garment vertex into the local coordinate frame of bone $b$ ($b\in B$). The garment is then reconstructed by linearly blending the corresponding inverse projections from all bones, weighted by skinning weights ${w_b}$, visualized in different colors for different bones.
  %We define a projection function $P_b$ for each bone $b\in B$. Then we express garment geometry $G$ as a linear blend of per-bone projections. \Kang{Explain g, wb, meaning of arrows.}
  }
  \label{fig:representation}
\end{figure}

As shown in Figure~\ref{fig:representation}, our method constructs the LoBoMap Blending representation in three steps: (1) defining a local coordinate frame for each bone, (2) {mapping} garment vertices into these frames (and defining the inverse mapping back to 3D space), and (3) reconstructing the garment by linearly blending the inverse mappings using skinning weights.

% \begin{wrapfigure}{r}{0.25\columnwidth}
%   \vspace{-0.8\baselineskip}
%   \centering
%   \includegraphics[0.24\columnwidth]{figs/draft/root_frame.pdf}
%   % \fbox{\rule{0pt}{3cm}\rule{0.18\textwidth}{0pt}} % height, width
%   %\caption{Inset (placeholder).}
%   \label{fig:hips-crotch-inset}
%   \vspace{-0.8\baselineskip}
% \end{wrapfigure}
% \begin{wrapfigure}{R}{0.35\columnwidth}
% \centering
% \includegraphics[width=0.35\columnwidth]{figs/draft/root_frame.pdf}
% \end{wrapfigure}
\paragraph{\textbf{Local Bone Frame Construction.}} 
Given a skeleton $B$ embedded in an avatar, we establish orthonormal frames for each bone $b \in B$ with a head joint located at $b^o$ and bone direction vector $\vec{b}$ of length $\ell_b=\|\vec{b}\|$.
Starting from the pelvis root bone, we initialize the root frame by orthogonalizing the hip bone direction with a crotch-reference vector to ensure stability across poses. For each bone $b$ with parent $b_{p}$, we construct a local frame $\left[\vec{b}^{x},\vec{b}^{y},\vec{b}^{z}\right]$, where:\\
\begin{itemize} [label=-, leftmargin=10pt]
    \item The \textbf{z-axis} is set to be the normalized bone direction $\vec b^z = \vec{b}/\|\vec{b}\|$;
    \item The \textbf{x-axis} is propagated from the parent frame via Gram-Schmidt orthogonalization $\vec b^x = \tilde b^x /\|\tilde b^x \|$, where $\tilde b^x =\vec b_{p}^x-(\vec b_{p}^x\cdot \vec b^z)\vec b^z$;
    \item The \textbf{y-axis} completes the right-handed system via cross product  $\vec b^y = \vec b^z \times \vec b^x$.
\end{itemize}
Note that {if $\|\tilde b^x\|<10^{-6}$ (i.e., $\vec {b}_{p}^x$ is nearly parallel to $\vec b^z$), we use $z$-axis from the parent frame as a fallback direction: $\vec b^x = \operatorname{sign}(-\vec b_{p}^x \cdot \vec b^z)\, \vec b_{p}^z$, where $\operatorname{sign}(0)$ is treated as $+1$.} %Finally, we complete a right-handed orthonormal basis by $\vec b_j^y = \vec b_j^z \times \vec b_j^x$, yielding the local bone frame.
%
%This hierarchical propagation ensures consistent frame orientation across kinematic chains, as visualized in Figure~\ref{fig:representation}.
This hierarchical propagation yields a coherent and consistent frame construction along kinematic chains, as visualized in Figure~\ref{fig:representation}.

\vskip -0.1in

\paragraph{\textbf{Local Bone Mapping and {Reconstruction}.}}
For a bone $b$ with frame $\left[\vec{b}^{x},\vec{b}^{y},\vec{b}^{z}\right]$ originating at $b^o$ and length $\ell_b$, we define: 
\begin{itemize} [label=-, leftmargin=10pt]
    \item \textbf{Mapping} $P_b(g)$, that maps a vertex $g\in\mathbb{R}^3$ to bone-local coordinates $(x_b, y_b, z_b)$:
  \[
  (x_b, y_b, z_b) \coloneqq P_b(g) = \frac{1}{\ell_b}\left(\langle g - b^o, \vec{b}^x \rangle, \langle g - b^o, \vec{b}^y \rangle, \langle g - b^o, \vec{b}^z \rangle\right)
  \]
  \item \textbf{{Reconstruction}} $P_b^{-1}(x_b, y_b, z_b)$, {i.e., the inverse mapping of $P_b$,} that reconstructs global position:
  \[
  g \coloneqq  P^{-1}_b(x_b,y_b,z_b) = b^o + \ell_b\left(x_b \vec{b}^x + y_b \vec{b}^y + z_b \vec{b}^z\right)
  \]
\end{itemize}
% - \textbf{Projection} $P_b(g)$, that maps a vertex $g\in\mathbb{R}^3$ to bone-local coordinates $(x_b, y_b, z_b)$:
%   \[
%   (x_b, y_b, z_b) \coloneqq P_b(g) = \frac{1}{\ell_b}\left(\langle g - b^o, \vec{b}^x \rangle, \langle g - b^o, \vec{b}^y \rangle, \langle g - b^o, \vec{b}^z \rangle\right)
%   \]
% - \textbf{Inverse projection} $P_b^{-1}(x_b, y_b, z_b)$, that reconstructs global position:
%   \[
%   g \coloneqq  P^{-1}_b(x_b,y_b,z_b) = b^o + \ell_b\left(x_b \vec{b}^x + y_b \vec{b}^y + z_b \vec{b}^z\right)
%   \]
%Here, $\langle\cdot,\cdot\rangle$ denotes the dot product. 
This formulation exhibits \textbf{equivariance}: keeping the local coordinates $(x,y,z)$ fixed and updating the bone frame yields coherent vertex motion under bone transformations (e.g., translation and rotation, as well as changes in bone length). Equivalently, $P_b(g)$ preserves a vertex’s relative placement in the local bone frame, while the decoded position naturally follows the skeletal motion.

\paragraph{\textbf{Linear Blending of {Reconstructions}.}}
We represent garment geometry as a linear blend of per-bone reconstructions. 
Given an avatar skeleton $B$, we define, for each garment vertex \(g\in\mathbb{R}^3\), its bone-local coordinates in bone $b$'s frame as
$ (x_b,y_b,z_b) \coloneqq P_b(g), \forall\, b\in {B}$.
The garment vertex $g$ is then represented as a blend of its {reconstructions}:
\[
g = \sum_{b \in B} w_b P_b^{-1}\left(x_b, y_b, z_b\right), \quad \text{where} \quad \sum_{b \in B} w_b = 1
\]
% We then express $g$ by blending the corresponding inverse projections:
% \begin{equation}
% g \coloneqq \sum_{b\in{B}} w_b\,P_b^{-1}(x_b,y_b,z_b),
% \qquad \text{with } \sum_{b\in{B}} w_b = 1.
% \end{equation}
{Here, $w_b$ is the blending weight for $b$, inherited and normalized from the source avatar's skinning weights of the nearest avatar-body vertex to $g$.}

\section{Garment Refitting} \label{sec: Garment Refitting}

Building upon the \emph{LoBoMap Blending} representation, we present the \emph{LoBoFit} algorithm for garment refitting. Given a source garment $G^s$ on avatar $A^s$ and a target avatar $A^t$ (sharing a similar skeletal topology and pose), LoBoFit generates a refitted garment $G^t$ for $A^t$ through initialization and optimization, as outlined in Figure~\ref{fig:overview}.

\subsection{Initialization via Mapping Transfer} \label{subsec: Ini}

%The refitting process begins with a pose-robust initialization. We transfer the source garment $G^s$ to the target avatar by reusing its pre-computed local coordinates $(x_{b^s}, y_{b^s}, z_{b^s}) \coloneqq P_{b^s}(g^s)$ for each bone $b^s \in B^s$ and each source garment vertex $g^s \in G^s$. These coordinates are decoded using the inverse projection $P_{b^t}^{-1}$ defined on the corresponding target bones $b^t \in B^t$.
The refitting process begins with a pose-robust initialization. We transfer the source garment $G^s$ to the target avatar by reusing its precomputed bone-local coordinates $(x_{b^s}, y_{b^s}, z_{b^s}) \coloneqq P_{b^s}(g^s)$ for each source vertex $g^s \in G^s$ and source bone $b^s \in B^s$. We then decode these coordinates via the inverse {map} $P^{-1}_{b^t}$ defined on the corresponding target bones $b^t \in B^t$.
The initial garment vertex $\hat{g}^t$ corresponding to $g^s$ is computed as:
\[
\hat{g}^t \coloneqq \sum_{b\in {B}} w_b^*\, P^{-1}_{b^t} \left(P_{b^s}(g^s)\right) = \sum_{b \in B} w_b^s P_{b^t}^{-1}(x_{b^s}, y_{b^s}, z_{b^s}),
\]
where the blending weights $w_b^*$ are initialized as $w_b^s$ that are inherited from the source garment.  Note that here we assume the source and target rigging bones use consistent bone indexing. That is, we introduce a common index set ${B}$: for each $b\in B$, $b^s\in {B}^s$ and $b^t\in {B}^t$ denote the corresponding source and target bones, respectively.

Our initialization leverages the \textbf{equivariance} of LoBoMap Blending, ensuring the initialized garment $\hat{G}^t$ is semantically attached to the target skeleton. As shown in Figure~\ref{fig:overview}, even in the same pose, source and target avatars may differ in bone-length proportions and local joint angles due to shape variation, and our method can provide a stable initialization with consistent bone attachment.

\subsection{Optimization Variables}\label{subsec: objectives}
As shown in Figure~\ref{fig:overview}, the initialized garment $\hat G^t$ follows the target skeleton $B^t$ but does not yet conform to the target body shape $A^t$.
Starting from $\hat G^t$,  we refine the target garment $G^t$ by optimizing two sets of variables defined in our \emph{LoBoMap Blending} representation: (i) \emph{Bone-Local Coordinate Residuals}  $\Delta_b \coloneqq (\Delta x_b,\Delta y_b,\Delta z_b)$, and (ii) \emph{Blending Weight Residuals} $\Delta w_b$, for each $b\in {B}$.
Specifically, for each garment vertex, the final position $g^t$ is defined as:
\begin{align}
g^t 
&= \sum_{b\in{B}} w_b^t \, P^{-1}_{b^t} \left(x_{b^t},y_{b^t},z_{b^t}\right) \\
&= \sum_{b\in{B}} w_b^t \, P^{-1}_{b^t} \left(x_{b^s}+\Delta x_b,\; y_{b^s}+\Delta y_b,\; z_{b^s}+\Delta z_b\right),
\end{align}
where $(x_{b^s},y_{b^s},z_{b^s})$ are the source bone-local coordinates transferred in initialization.

Note that while the source-transferred blending weights $w_b^s$ are usually effective for body-tight garments, ambiguous bone assignments can arise, especially between the two legs (near the crotch region as shown in Figure \ref{fig:overview}). In such cases, fixing $w_b^s$ may hinder convergence and lead to visible artifacts.
% The blending weights transferred from the source $w_{b^s}$ typically work well for body-tight garments, but keeping them fixed can cause noticeable artifacts for loose garments. 
We therefore jointly optimize blending weight residuals and define the target blending weights $w_b^t$  by normalizing the perturbed weights:
\begin{equation}
w_b^t \coloneqq \frac{w_b^s+\Delta w_b}{\sum_{k\in {B}} \left(w_k^s+\Delta w_k\right)}.
\end{equation}
% Moreover, to keep the weight residuals within a controllable range, we do not optimize $\Delta w_b$ directly. Instead, we optimize an unconstrained variable $\Delta\hat w_b$ and compute
% $\Delta w_b \coloneqq \gamma\,\tanh \bigl(\Delta\hat w_b\bigr) $ ,
% where $\tanh(\cdot)\in[-1,1]$ constrains $\Delta w_b(g)\in[-\gamma,\gamma]$. We set $\gamma = 0.1$ in our experiments.

\subsection{Loss Functions} \label{subsec: losses}
%To remove garment–body interpenetrations while preserving garment shape, user-specified garment-body fit style, and fine-scale wrinkles, we minimize a composite loss function: 
To guide the optimization process towards a high-quality refitting result, we minimize a composite loss function:
\begin{equation}
\mathcal {L} = \lambda \mathcal {L}_{\text {contact}} + \mathcal {L}_{\text{pres}} + \mathcal {L}_{\text{reg}},
\end{equation}
where $\lambda = 10$ controls the relative importance of the contact term. 
$\mathcal {L}_{\text {contact}} $ penalizes garment–body interpenetrations and enforces user-specified contact constraints (garment-body fit).
$\mathcal {L}_{\text{pres}} $ preserves garment shape and wrinkle details by encouraging consistency between the target $G^t$ and the source $G^s$. 
$\mathcal {L}_{\text{reg}} $ regularizes the optimization variables to maintain stable garment–body associations and improve convergence.
This formulation balances collision resolution, detail preservation, and optimization stability.
%where $\lambda$ controls the relative importance of the contact term and we use $\lambda=10$ for all experiments.

%\subsubsection*{\textbf{Contact Loss $\mathcal{L}_{\text{contact}}$}}
% \textbf{Contact Loss $\mathcal{L}_{\text{contact}}$.}
%This term ensures the garment conforms to the target body without penetrations while maintaining the source fit style.
\begin{figure*}[!htbp]
    \centering
    \includegraphics[width=0.95\textwidth]{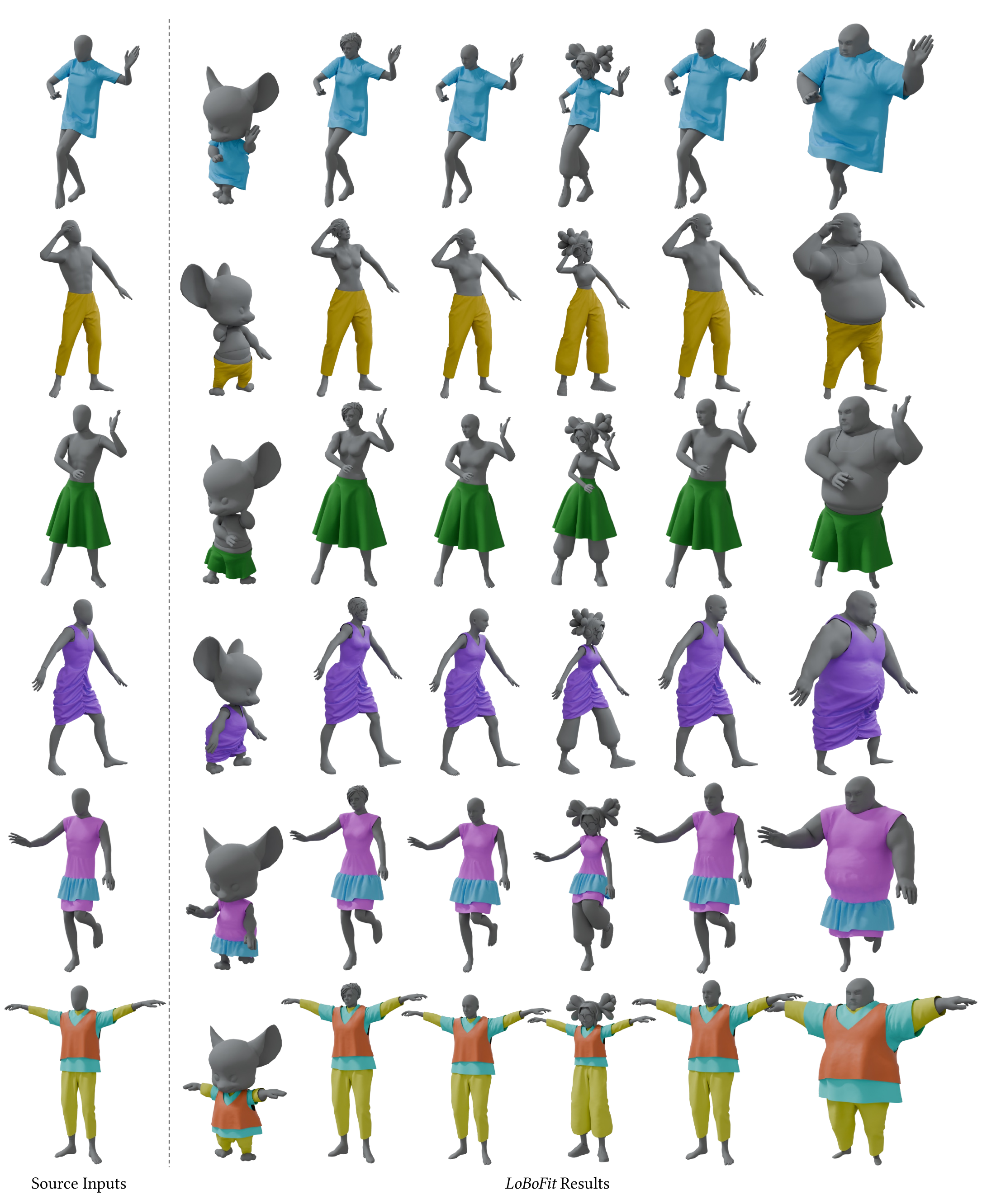}
    \vskip -0.1in
    \caption{\textbf{Results.} Given garments designed for a source avatar, we use \emph{LoBoFit} to refit them to target avatars while preserving design features and fine wrinkle details. Please zoom in to see the details more clearly.}
    \label{fig:results}
\end{figure*}

\begin{figure*}[!bhtp]
    \centering
    \includegraphics[width=\textwidth]{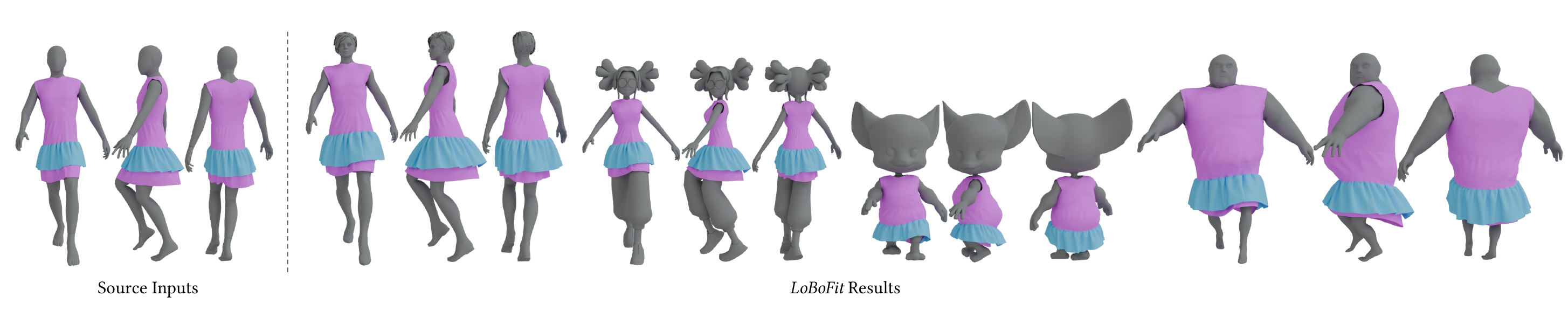}
    \vskip -0.12in
    \caption{\textbf{Extension to dynamic garments.} 
    We apply \emph{LoBoFit} frame-by-frame to refit dynamic garments while preserving time-varying wrinkles. We show a challenging two-layer dress sequence to highlight the effectiveness and robustness of \emph{LoBoFit}. Sequential results are in the supplemental video.
    }
    \label{fig:dynamic}
\end{figure*}

\paragraph{\textbf{Contact Loss $\mathcal{L}_{\text{contact}}$.}}
This term promotes body-conforming, penetration-penalized fitting while preserving the source fit style.
%\Zhang{To this end, collision handling is formulated as a soft penalty based on nearest-point queries between garment vertices and the target body. We initialize contact pairs using a bone-guided proximity strategy and update them during optimization (see supplemental material for details).}
{To this end, collision handling is formulated as a soft penalty based on nearest-point queries between garment vertices and the target body. Contact pairs are initialized using a bone-guided proximity strategy, which is used only at the beginning of optimization, and are subsequently updated every 2000 iterations using an approximate nearest-face search based on KNN over target-body face centroids once the garment moves close to the target body (see supplemental material for details).}
We define:
\[
\mathcal{L}_{\text{contact}} = \mathcal{L}_{\text{sep}} + \mu_t \mathcal{L}_{\text{tight}}
\]
where $\mu_t$ controls the strength of garment-body fit style matching, and we set $\mu_t=10$ in all experiments. The loss terms consist of:
\begin{itemize} [label=-, leftmargin=10pt]
    \item \textbf{Separation Loss} $\mathcal{L}_{\text{sep}}$, that penalizes garment-body interpenetrations beyond a predefined threshold $\epsilon $:
        \[
    \mathcal{L}_{\text{sep}} = \sum_{(g^t, a^t)} \omega(a^t) \left[ \max(0, \epsilon - d_g^t) \right]^2,
    \]
    where $d_g^t$ is the signed distance from garment vertex $g^t$ to its nearest point $a^t$ on the target body, and $\omega(a^t)$ is a normalized weight proportional to the area of the target-body face containing $a^t$. More specifically, $d_g^t = n_a^\top(g^t-a^t)$,
    where $n_a$ is the outward surface normal at $a^t$.

    \item \textbf{Tightness Loss} $\mathcal{L}_{\text{tight}}$, that preserves fit-style consistency by matching signed distances within a user-specified fit-control region $\mathcal{F}$:
    \[
    \mathcal{L}_{\text{tight}} = \sum_{g \in \mathcal{F}} (d_g^t - d_g^s)^2.
    \]
    The region $\mathcal{F}$ specifies where the refitted garment should inherit the source fit style (e.g., "Upper trunk", "Waist", or "Upper trunk \& waist"), which is automatically partitioned based on semantic bone associations from \emph{LoBoMap Blending}. 
   Please refer to the supplemental material for more details.
\end{itemize}
\paragraph{\textbf{Preservation Loss $\mathcal{L}_{\text{pres}}$.}}
This term maintains the source garment's geometric details and is defined as:
\[
\mathcal{L}_{\text{pres}} = \mu_l \mathcal{L}_{\text{lap}} + \mathcal{L}_{\text{bend}} + \mathcal{L}_{\text{curv}},
\]
where $\mu_{l}$ is a weighting coefficient, and we set $\mu_{l} = 0.5$ in all experiments. The loss terms consist of:
\begin{itemize}[label=-, leftmargin=10pt]
    \item 
    \textbf{Laplacian Term} $\mathcal{L}_{\text{lap}}$, {that preserves local differential structure and high-frequency wrinkles} via normalized Laplacian coordinates:
    %that is  crucial for preserving high-frequency wrinkles 
    \[
    \mathcal{L}_{\text{lap}} = \sum_i \| \tilde{\delta}(g_i^t) - \tilde{\delta}(g_i^s) \|_2^2,
    \]
    where $\tilde{\delta}(g_i) = \delta(g_i) / (\sum_k \|\delta(g_k)\|_2^2)^{1/2}$ normalizes the Laplacian coordinates $\delta(g_i)$ to account for differences in overall Laplacian magnitude, with $\delta(g_i) = \sum_{j\in N(i)} c_{ij}\,(g_j-g_i)$, where $N(i)$ is the 1-ring neighborhood of vertex $i$ on $G^s$ and $c_{ij}$ are precomputed cotangent weights~\cite{meyer2003discrete}. %Accordingly, $\delta(g_i^s)$ applies this operator to the source vertex $g_i^s$, while $\delta(g_i^t)$ applies the same operator (with the same $N(i)$ and $c_{ij}$) to the corresponding target vertex $g_i^t$.
    
    \item 
    \textbf{Bending Term} $\mathcal{L}_{\text{bend}}$, {an empirically motivated regularizer} that improves global smoothness and suppress spurious sharp creases by {encouraging consistency of} dihedral angles $\theta$  between garment faces $f_i$ and $f_j$:
    \[
    \mathcal{L}_{\text{bend}} = \sum_{(i,j)} \omega(e_{ij}^s) \left( \theta(f_i^t, f_j^t) - \theta(f_i^s, f_j^s) \right)^2.
    \]
    where $f^s$ and $f^t$ are the corresponding faces in the source and target garments, and $\omega(e^s_{ij})$ is a normalized weight predefined by the length of the edge $e^s_{ij}$ shared by $f^s_i$ and $f^s_j$ based on the source garment.
    
    \item 
    {\textbf{Curvature Term} $\mathcal{L}_{\text{curv}}$,} that preserves garment silhouette and style by matching boundary edge curvatures using the cosine angle between consecutive boundary edges:
    {\[
    \mathcal{L}_{\text{curv}} = \sum_i \omega(\ell_i^s) \left( \hat{e}_i^t \cdot \hat{e}_{i+1}^t - \hat{e}_i^s \cdot \hat{e}_{i+1}^s \right)^2,
    \]}
    % \[
    % \mathcal{L}_{\text{curv}} = \sum_i \omega(\ell_i^s) (e_i^t \cdot e_{i+1}^t - e_i^s \cdot e_{i+1}^s)^2.
    % \]
    where {$\hat{e}_i = e_i / \|e_i\|$} and $e_i^s,e_{i+1}^s$ (respectively $e_i^t,e_{i+1}^t$) are adjacent boundary edges in the source (respectively target) garment. $\omega(\ell_i^s)$ is a normalized weight determined by the lengths of the corresponding boundary segments $\ell_i^s=\|e_i^s\|$ in the source garment.
    %\Kang{[KKZ: $\omega(\ell_i^s)$  needs to be more specific.]}
\end{itemize}

\paragraph{\textbf{Regularization $\mathcal{L}_{\text{reg}}$.}}
This term stabilizes optimization by constraining residual changes:
\[
\mathcal{L}_{\text{reg}} = \mathcal{L}_{\Delta z} + \mu_w \mathcal{L}_{\Delta w}.
\]
Here, $\mathcal{L}_{\Delta z} = \sum_z \sum_{b \in B} (\Delta z_b)^2$ penalizes large changes along bone axes to preserve source-consistent bone attachment,
%encouraging the garment to remain consistently attached to the corresponding bones as in the source. 
and $\mathcal{L}_{\Delta w} = \sum_w \sum_{b \in B} (\Delta w_b)^2$ regularizes blending weight residuals. $\mu_w$ is a weighting coefficient, and we set $\mu_w = 0.01$ in all experiments.

\subsection{Optimization Strategy} \label{subsec: optimization}

Given the initialization $\hat G^t$, we optimize for the refitted garment $G^t$ using AdamW~\cite{loshchilov2019adamw} with a learning rate of $5\times10^{-3}$ and \texttt{amsgrad} enabled. 
We iteratively optimize the mapping residuals $(\Delta x_b,\Delta y_b,\Delta z_b)$ and the weight residuals $\Delta w_b$, starting from all residuals initialized to $0$.
% the unconstrained weight residuals $\Delta\hat w_b$ (from which $\Delta w_b$ is computed), starting from all residuals initialized to zero. 
Moreover, to ensure stable convergence, we apply a soft constraint to weight residuals by optimizing an unconstrained variable $\Delta \hat{w}_b$ and computing:
\[
\Delta w_b = \gamma \tanh(\Delta \hat{w}_b), \quad \gamma = 0.1.
\]
This confines weight adjustments to the range $[-0.1, 0.1]$, preventing overly drastic changes to bone influence weights.
%to keep the weight residuals within a controllable range, we do not optimize $\Delta w_b$ directly. Instead, we optimize an unconstrained variable $\Delta\hat w_b$ and compute $\Delta w_b\coloneqq\gamma\,\tanh \bigl(\Delta\hat w_b\bigr)$, where $\tanh(\cdot)\in[-1,1]$ constrains $\Delta w_b(g)\in[-\gamma,\gamma]$. We set $\gamma=0.1$ in our experiments.
In practice, a plausible refitted garment that fits the target avatar $A^t$ is typically obtained after $\sim$8,000 iterations. We update the garment-body contact pairs $(g^t,a^t)$ via a standard nearest-neighbor search every 2,000 iterations.

For high-resolution garments with complex geometry and fine wrinkles, directly optimizing at the target resolution often yields good results, but can be less reliable in challenging cases—occasionally getting trapped in suboptimal local minima and producing artifacts—and is also computationally expensive. We therefore adopt a hierarchical coarse-to-fine optimization procedure inspired by~\cite{zhang2024neural}, which improves both robustness and efficiency. 

We provide more implementation details and practical considerations in the supplemental material.

\begin{figure*}[!htbp]
    \centering
    \includegraphics[width=\textwidth]{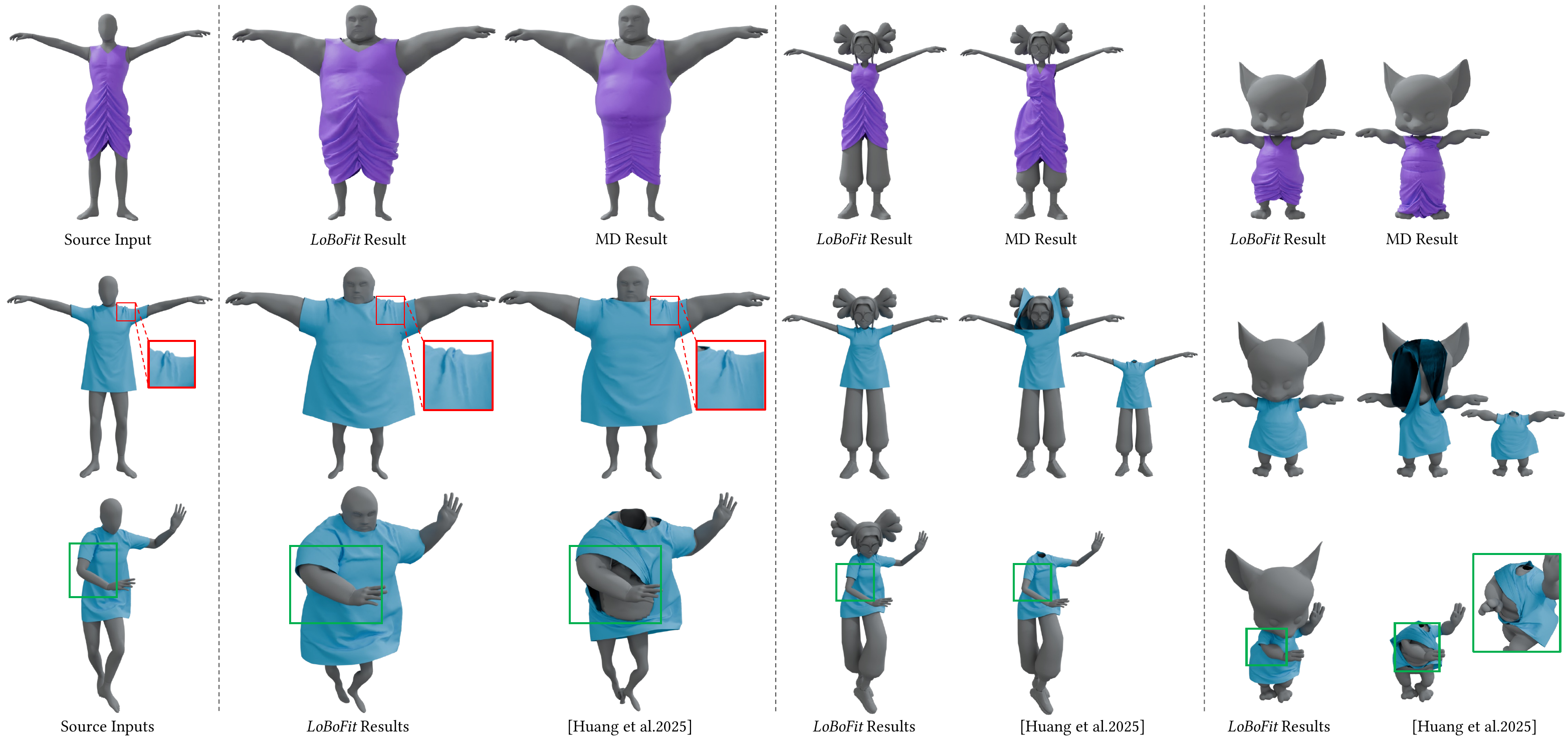}
    \vskip -0.1in
    \caption{\textbf{Comparisons.} 
    1st row: Compared with re-draping the garment on the target avatar using the fitting tool in MD, \emph{LoBoFit} better maintains garment–body correspondences while preserving design features and fine-scale wrinkles. 2nd and 3rd rows: Compared with IFGR~\cite{huang2025intersection}, \emph{LoBoFit} is more robust under non-canonical poses and more faithfully preserves fine-scale details.
    }
    %We compare \emph{LoBoFit} with IFGR~\cite{huang2025intersection}. \emph{LoBoFit} is more robust across diverse avatars and poses, and more faithfully preserves fine-scale wrinkle details.}
    \label{fig:comparison}
\end{figure*}

\begin{figure*}[!htbp]
    \centering
    \includegraphics[width=\textwidth]{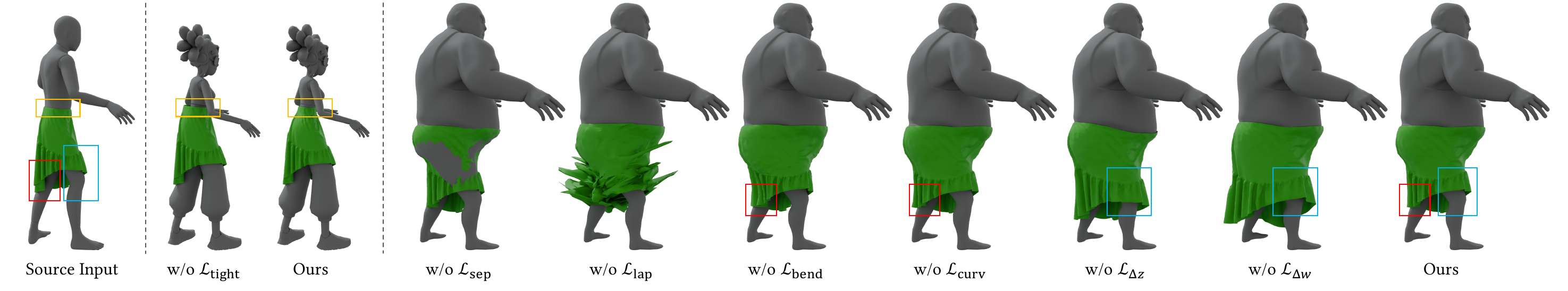}
    \vskip -0.1in
    \caption{\textbf{Ablation on Loss Terms.} Removing $\mathcal{L}_{\text{tight}}$ breaks fit-style consistency under proportion changes (yellow boxes); removing $\mathcal{L}_{\text{sep}}$ leads to interpenetration. 
    Disabling $\mathcal{L}_{\text{lap}}$ degrades wrinkle/detail transfer; dropping $\mathcal{L}_{\text{bend}}$/$\mathcal{L}_{\text{curv}}$ introduces unnatural hem folds (red boxes), and removing remaining regularizers ($\mathcal{L}_{\Delta z}$ and $\mathcal{L}_{\Delta w}$) causes drift and misplacement (blue boxes). Full objective $\mathcal{L}$ preserves fit, style, and fine wrinkles across diverse body shapes.
}
    \label{fig:ablat}
\end{figure*}

\section{Experiments and Results}
We evaluate \emph{LoBoFit} by retargeting garments to diverse target avatars and poses while preserving garment design features and fine-scale wrinkle details, demonstrating both effectiveness and robustness (Section \ref{subsec:results}). To highlight the benefits of our geometry representation, \emph{LoBoMap Blending}, and our refitting method, \emph{LoBoFit}, we compare with an existing garment authoring platform, as well as IFGR, a state-of-the-art garment refitting approach (Section \ref{subsec: compare}). We further perform ablation studies to assess the contribution of each key component of our method (Section \ref{subsec:eval}).

%\subsection{Avatars and Garments}
\subsection{Results}\label{subsec:results}
\emph{LoBoFit} successfully refits single- and multi-layer garments to avatars with large shape and topological variations, as shown in Figure~\ref{fig:results}. We created and rigged target avatars using Mixamo, covering a diverse range of body shapes. Source garments were authored in Marvelous Designer (MD) with a particle distance of 10mm, resulting in single-layer garments with 10k–30k vertices to capture fine wrinkles.
Using our coarse-to-fine optimization strategy, refitting a single-layer garment takes approximately 1 {minute} on {an} NVIDIA GeForce RTX 5090.
%The runtime scales primarily with the number of garment vertices and is insensitive to the complexity of the avatar body model. 
We provide data details in the supplemental material. 

%In Figure~\ref{fig:results}, we demonstrate the effectiveness of our garment refitting method, \emph{LoBoFit}, by retargeting six high-resolution garment models (including two multi-layer garments) from a source mannequin to six different target avatars under six corresponding poses. 

%\Kang{In detail, we create and rig the target avatar using Mixamo, which covers a diverse range of body shapes and mesh topologies. We author the source garments in Marvelous Designer (MD) with a particle distance of 10mm;} at this setting, a single-layer garment typically contains {10k--30k} vertices, allowing us to represent fine-grained wrinkle details. We provide data details in the supplemental material. 
%Using our coarse-to-fine optimization strategy, refitting a single-layer garment takes approximately 2 minutes on a \Kang{NVIDIA} GeForce RTX 5090. 
%The runtime scales primarily with the number of garment vertices and is insensitive to the complexity of the avatar body model. 

%\paragraph{Garment Refitting}

%\paragraph{Extension to Multi-layer Garments}
\textbf{Extension to Multi-layer Garments.}
\emph{LoBoFit}  naturally extends to multi-layer garments (Figure~\ref{fig:results}, rows 5-6). We refit layers sequentially from inner to outer, treating the optimized inner layer as part of the target avatar's collision geometry when fitting the outer layer.
%Figure~\ref{fig:results} shows two multi-layer garment results (rows 5 and 6), demonstrating that \emph{LoBoFit} naturally extends to multi-layer refitting. We refit the layers sequentially, treating the refitted inner layer as part of the target avatar geometry when optimizing the outer layer. 
Please refer to the supplemental material for details of the optimization strategy.

%\paragraph{Extension to Dynamic Garments}
\textbf{Extension to Dynamic Garments.}
Since LoBoFit is not restricted to a canonical pose~\cite{chen2025dress,huang2025intersection}, it can process dynamic sequences frame-by-frame. Figure~\ref{fig:dynamic} shows a challenging two-layer dress sequence where time-varying wrinkles are retained, demonstrating \emph{LoBoFit}'s effectiveness and robustness for dynamic garment retargeting. 
%Sequential results are in the supplemental video.
%Since \emph{LoBoFit} does not restrict refitting to a canonical pose~\cite{chen2025dress,huang2025intersection}, it \Kang{can} naturally extend to dynamic garment refitting\Kang{: g}iven a sequence of source garments simulated over a motion sequence on the source avatar, \emph{LoBoFit} refits the garment frame-by-frame to the target avatar under the corresponding poses, while faithfully retaining the fine-scale, time-varying wrinkle dynamics. Figure~\ref{fig:dynamic} shows a challenging two-layer garment example of this dynamic extension to highlight the effectiveness and robustness of \emph{LoBoFit}.  
%We demonstrate more dynamic garment results in the supplemental material.

\subsection{Baseline Comparisons} \label{subsec: compare}
We compare against the refitting tool in Marvelous Designer (MD), which requires users to manually build a tight cage around the source and target avatars (often in a canonical pose), then re-simulate the same sewing {pattern} and re-drape the garment on the target, without explicitly preserving the original fit style. In contrast, \emph{LoBoFit} better maintains garment–body correspondences while preserving design features and fine-scale wrinkles (1st row in Figure~\ref{fig:comparison}).
We also compare against IFGR~\cite{huang2025intersection}, which follows a {skeletal embedding and inflation} pipeline optimized with an IPC-based solver~\cite{li2020incremental}. We conduct this comparison on a T-shirt example (with 16{,}058 vertices), using the same source avatar in two different poses. For a fair comparison, we use the same fit-control regions $\mathcal{F}$ for both \emph{LoBoFit} and IFGR. These regions are automatically constructed by our \emph{LoBoMap} partitioning from the user’s semantic input {labels (e.g, "Waist" and "Upper trunk")}, whereas the original IFGR requires manually specifying fit-control regions. We then refit the source garment to three stylized target avatars in the corresponding poses. 

As illustrated in Figure~\ref{fig:comparison} (2nd and 3rd rows), \emph{LoBoFit} demonstrates three key advantages:
 \textbf{(1) Robust Initialization:} IFGR relies on hard bone assignments determined by nearest-neighbor searches. Under non-canonical poses, this can lead to semantically incorrect attachments. For example, in Figure~\ref{fig:comparison}, the mouse avatar's ears are incorrectly assigned to the shoulder bone in the lateral-raise pose, causing garment misplacement during the inflation step. Even when excluding the head to rule out misalignment, failures persist when arm bones become closer to the abdominal regions than the spine (green boxes). In contrast, LoBoFit's initialization, built on the equivariant LoBoMap Blending representation, provides pose-consistent semantic attachment by reusing source local coordinates, yielding a stable starting point for optimization.
 \textbf{(2) Superior Detail Preservation:} IFGR preserves details by applying a global scale factor derived from avatar sizes, following~\cite{araujo2023slippage}.
 However, wrinkle characteristics are influenced by local metric distortions (stretching/compression) that vary spatially. A single global scaling cannot fully capture this, sometimes over-amplifying details in certain regions (red box, Figure~\ref{fig:comparison}). LoBoFit operates natively on the target avatar's scale and preserves details by matching normalized Laplacian coordinates between source and target garments. This differential constraint is less sensitive to global size differences while responsive to local geometric changes, producing wrinkle patterns that more closely resemble the source.
  \textbf{(3) Efficiency:} IFGR's IPC-based inflation is computationally expensive ($\sim 30$ minutes for a 16k-vertex garment) and sensitive to mesh imperfections. LoBoMap Blending decomposes the problem into bone-local subproblems, leading to a better-conditioned optimization. LoBoFit converges to a high-quality result in about 1 minute for the similar garment, demonstrating significantly higher efficiency.

{In the supplemental material, we further compare with IFGR using cycle-consistency on the lateral raised pose. \emph{LoBoFit} achieves lower error, indicating better preservation of source features. We also report self-intersection results: IFGR guarantees collision-free outputs via IPC, while our method shows only small residual intersections. These results support the effectiveness of our method.}

%We include further analysis and additional comparisons in the supplemental material.

% Moreover, we conduct a quantitative comparison with IFGR via a cycle consistency evaluation by reversing the retargeting from the target back to the source, i.e., $G^s \rightarrow G^t \rightarrow \tilde{G}^s$, where $\tilde{G}^s$ should conform to the source avatar $A^s$. \emph{LoBoFit} achieves a lower error, indicating better preservation of source features than IFGR. More details and future comparisons are in the supplementary material.
%We provide further comparisons with IFGR in the supplementary material.
% \begin{table}[h]
%   \centering
%   \caption{\textbf{Cycle consistency evaluation.} 
%   %We report the mean per-vertex distance between the original garment $G^s$ and the reconstructed garment $\tilde{G}^s$,$\text{mean}\left(\|g^s-\tilde{g}^s\|_2\right)$.
%   We report the mean per-vertex distance between the original garment $G^s$ and the reconstructed garment $\tilde{G}^s$: $\mathrm{mean}\left(\|g^s-\tilde{g}^s\|_2\right)$. \emph{LoBoFit} yields lower error.
% } 
%   \label{tab:cyc_eval}
%   \vskip -0.1in
%   \begin{tabular}{|c||c|c|c|}
%   \hline
%    & Mousey & Ortiz & Michelle \\
%    \hline
%    % $\mathcal{F}$ & "All" & "All" & "Upper trunk" \\
%    % \hline
%    \emph{LoBoFit} & \textbf{2.41e-2} & \textbf{1.28e-2} & \textbf{1.82e-2} \\
%    \hline
%    IFGR & 3.61e-2 & 2.36e-2 & 2.76e-2 \\
%   \hline
%   \end{tabular}
% \end{table}
%
\begin{figure}[!t]
    \centering
    \includegraphics[width=\columnwidth]{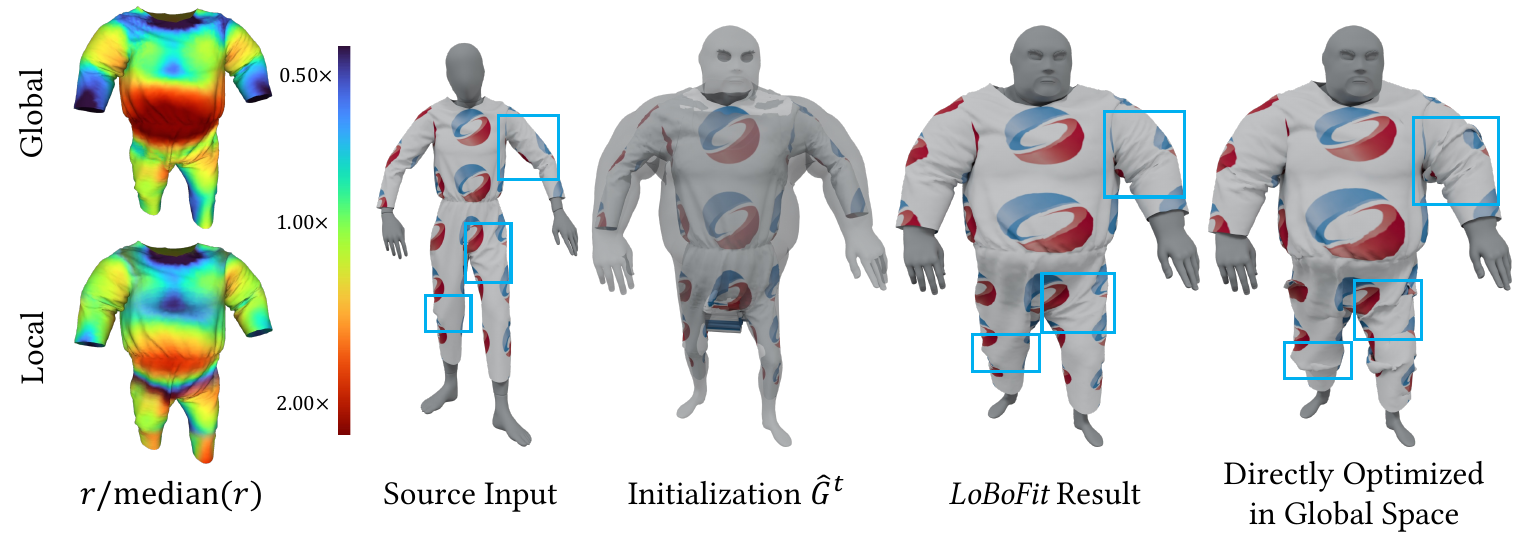}
    \vskip -0.1in
    \caption{
    \textbf{Effect of \emph{LoBoMap Blending}.} 
    %\textbf{Benefits of \emph{LoBoProj Blending}.} 
    It provides a pose-robust, coherently placed initialization $\hat{G}^t$, and yields a better-conditioned optimization by expressing deformations in bone-local subspaces (tighter displacement distribution than global space), enabling faster and more stable convergence than direct global vertex-offset optimization.
    }
    \label{fig:eff_loboproj}
\end{figure}

\begin{figure}[t]
  \centering
  %\fbox{\rule{0pt}{4cm}\rule{0.98\columnwidth}{0pt}} % height=4cm, width=0.8\linewidth
  \includegraphics[width=\columnwidth]{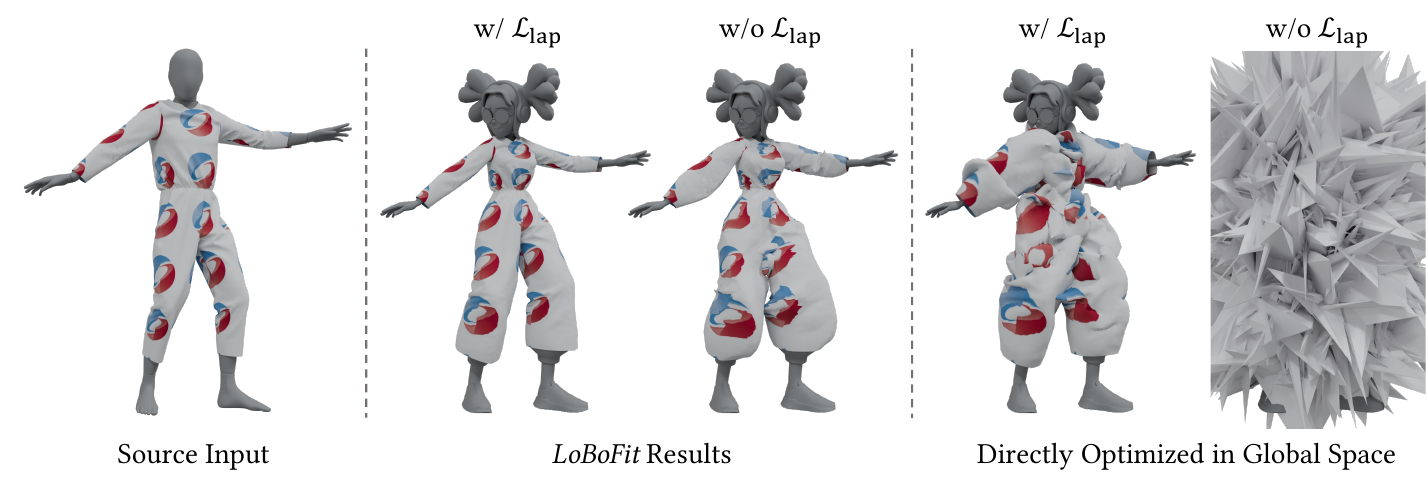}
  \vskip -0.1in
  \caption{
  %\textbf{Effect of \emph{LoBoProj Blending}.}
  \textbf{Conditioning comparison (\emph{LoBoFit} vs. Directly Optimized in Global)} 
  With $\mathcal{L}_{\text{lap}}$, our LoBoMap-based optimization yields clean results, whereas direct global offset optimization is prone to poor minima and artifacts. Without $\mathcal{L}_{\text{lap}}$, our bone-local \emph{LoBoMap Blending} formulation largely preserves low-frequency shape but loses fine wrinkles, while global offsets become unstable and break down, further evidencing the improved conditioning enabled by \emph{LoBoMap Blending}.
  }
  \label{fig:betten_condition}
\end{figure}
\subsection{Evaluations} \label{subsec:eval}
%\paragraph{Ablation Study on Loss Terms.}
\emph{Ablation Study on Loss Terms.}
We conduct an ablation study of our loss terms using the same source garment, refitted to two target avatars with representative body shapes (Figure~\ref{fig:ablat}). Removing $\mathcal{L}_{\text{tight}}$ breaks fit-style consistency under proportion changes: the garment remains close to the source and fails to adapt to the target waist-to-hip ratio (yellow boxes). Removing $\mathcal{L}_{\text{sep}}$ leads to visible garment--body interpenetrations. Disabling $\mathcal{L}_{\text{lap}}$, our key mesh-quality term, severely degrades detail transfer, producing distorted wrinkles and artifacts. Dropping the geometric regularizers $\mathcal{L}_{\text{bend}}$ and $\mathcal{L}_{\text{curv}}$ introduces unnatural folds, especially along hem boundaries (red boxes). Removing the remaining regularizers ($\mathcal{L}_{\Delta z}$ and $\mathcal{L}_{\Delta w}$) causes additional drift in garment--body placement and overall shape (blue boxes). In contrast, the full objective $\mathcal{L}$ enables \emph{LoBoFit} to robustly refit across diverse body shapes while preserving fit style, overall shape, and fine-scale wrinkles.

\begin{figure}[!htbp]
    \centering
    \includegraphics[width=\columnwidth]{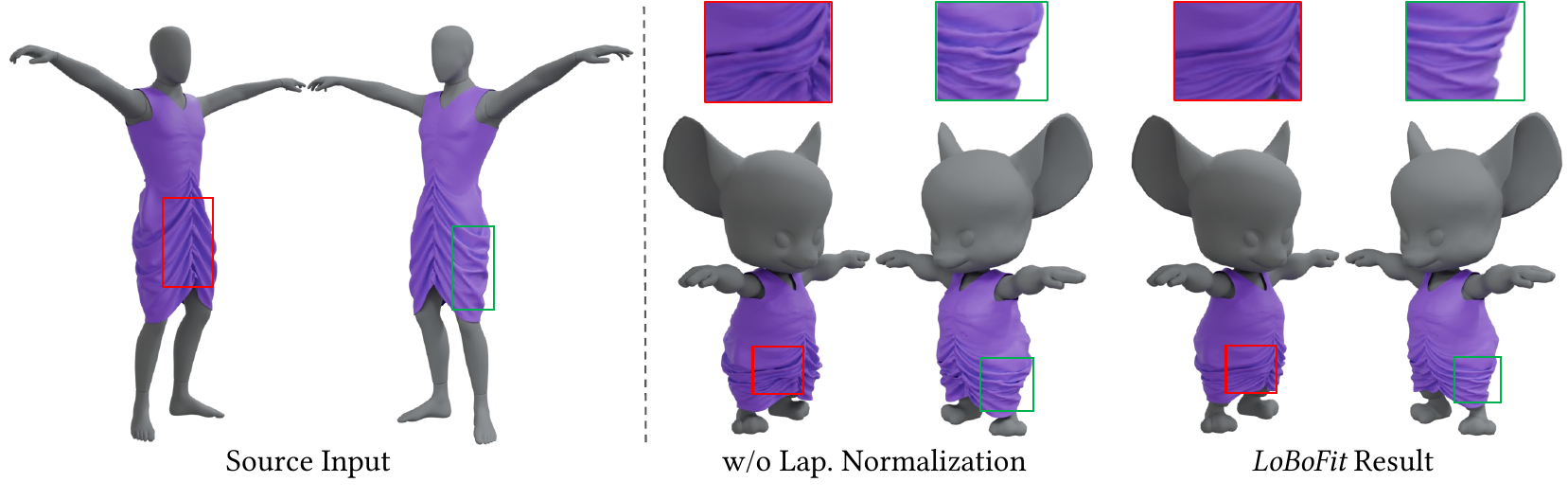}
    \vskip -0.1in
    \caption{\textbf{Effect of Laplacian Normalization.} %\emph{LoBoFit} with Laplacian feature normalisation better preserves source-consistent fine-scale wrinkles (ours), whereas \textbf{\textbackslash wo} normalization squeezes details and introduces artifacts.
    Using an unnormalized Laplacian term (w/o Lap. Normalization) on a much smaller target (mouse) compresses wrinkles and introduces self-intersections and artifacts, whereas Laplacian normalization compensates for global detail-scale differences and preserves plausible fine-scale wrinkles.
    }
    \label{fig:lapnorm}
\end{figure}

\emph{Effect of LoBoMap Blending.}
Figure~\ref{fig:eff_loboproj} highlights two key benefits of our \emph{LoBoMap Blending} representation.
\textbf{(1) Pose-robust initialization.} Running the full \emph{LoBoFit} pipeline, we visualize the initialization $\hat{G}^t$ and the final result, showing that \emph{LoBoMap Blending} provides a reliable, coherently placed starting point.
\textbf{(2) Better-conditioned optimization.} \emph{LoBoMap} decomposes deformation into bone-local subproblems, yielding a more compact displacement distribution. Visualizing $r/\mathrm{median}(r)$ with $r=\lVert \Delta\mathbf{x}\rVert$, global-space displacements are widely spread, whereas bone-local displacements cluster tightly (each offset $\Delta\mathbf{x}$ in its dominant-bone frame). Accordingly, \emph{LoBoFit} converges in $\sim$10k iterations (coarse+fine), while direct global offset optimization is slower and more prone to poor minima, often artifact-prone even after $>40$k iterations. This conditioning advantage is further evidenced by removing $\mathcal{L}_{\text{lap}}$: bone-local optimization remains stable, while global offsets become unstable and break down (see Figure \ref{fig:betten_condition}).
\emph{Effect of Laplacian Normalization.}
Figure~\ref{fig:lapnorm} illustrates the effect of Laplacian feature normalization. When computing $\mathcal{L}_{\text{lap}}$, we compare source and target Laplacian coordinates using either the raw features $\delta(g_i)$ or the normalized features $\tilde{\delta}(g_i)$. For the mouse avatar, which is much smaller than the source, using unnormalized $\delta(g_i)$ compresses wrinkle detail, causing self-intersections and visible artifacts. Normalizing the Laplacian features removes differences in overall Laplacian magnitude (i.e., global detail scale) between source and target, allowing \emph{LoBoFit} to adapt to the target scale while preserving plausible fine-scale wrinkles.

\section{Limitations and Future Work}
\begin{figure}[!htbp]
    \centering
    \includegraphics[width=0.6\columnwidth]{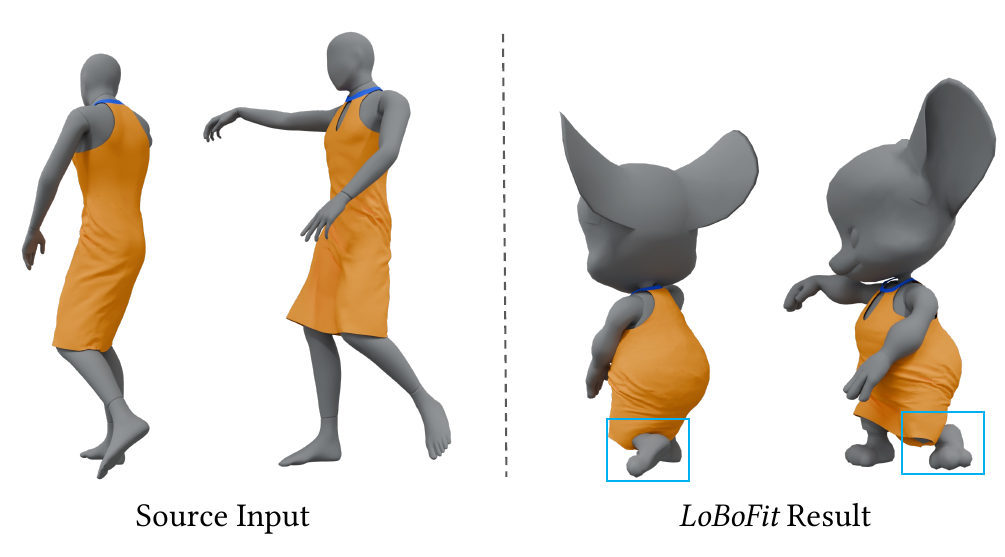}
    \vskip -0.1in
    \caption{\textbf{Failure Case.} \emph{LoBoFit} fails to eliminate all garment–body penetrations when feature preservation and intersection resolution conflicts.}
    \label{fig:failureCase}
\end{figure}

Our method has several limitations. 
First, {as a geometry-based method rather than a universal, physics-complete solution, \emph{LoBoFit} may leave residual garment-body penetrations when feature preservation conflicts with intersection resolution (Figure~\ref{fig:failureCase}).}
Second, our frame-by-frame dynamic refitting is purely geometric, which can cause temporal flickering; while temporal regularization may help, we leave physics-aware modeling for future work. 
{Third, our method builds upon an LBS-based representation, which may exhibit known artifacts such as local volume collapse or limited expressiveness under extreme deformations or large shape discrepancies.}
Finally, we do not explicitly model self-collision, {so residual self-intersections may occur, }especially when local regions are strongly compressed during retargeting.
A promising direction is physics-based refinement (e.g., collision handling and sewing-pattern optimization), with \emph{LoBoFit} providing a strong initialization for remeshing, parameter inference, and pattern adjustment.

\begin{acks}
We thank the anonymous reviewers for their valuable feedback and constructive suggestions, which helped improve this work. We gratefully acknowledge Mixamo for providing publicly available avatars and motion data used in our experiments. We also thank Marvelous Designer, whose membership-accessible garment library served as a starting point for several of the clothing assets used in this work.
This work was partially supported by the National Science Fund of China (No. 62472223 and No. 62322207). 
\end{acks}


%\clearpage
\bibliographystyle{ACM-Reference-Format}
\bibliography{main_bib}

% \appendix
% \input{8_figures}

\appendix
\begin{figure}[h]
  \centering
  %\fbox{\rule{0pt}{4cm}\rule{0.98\columnwidth}{0pt}} % height=4cm, width=0.8\linewidth
  \includegraphics[width=\columnwidth]{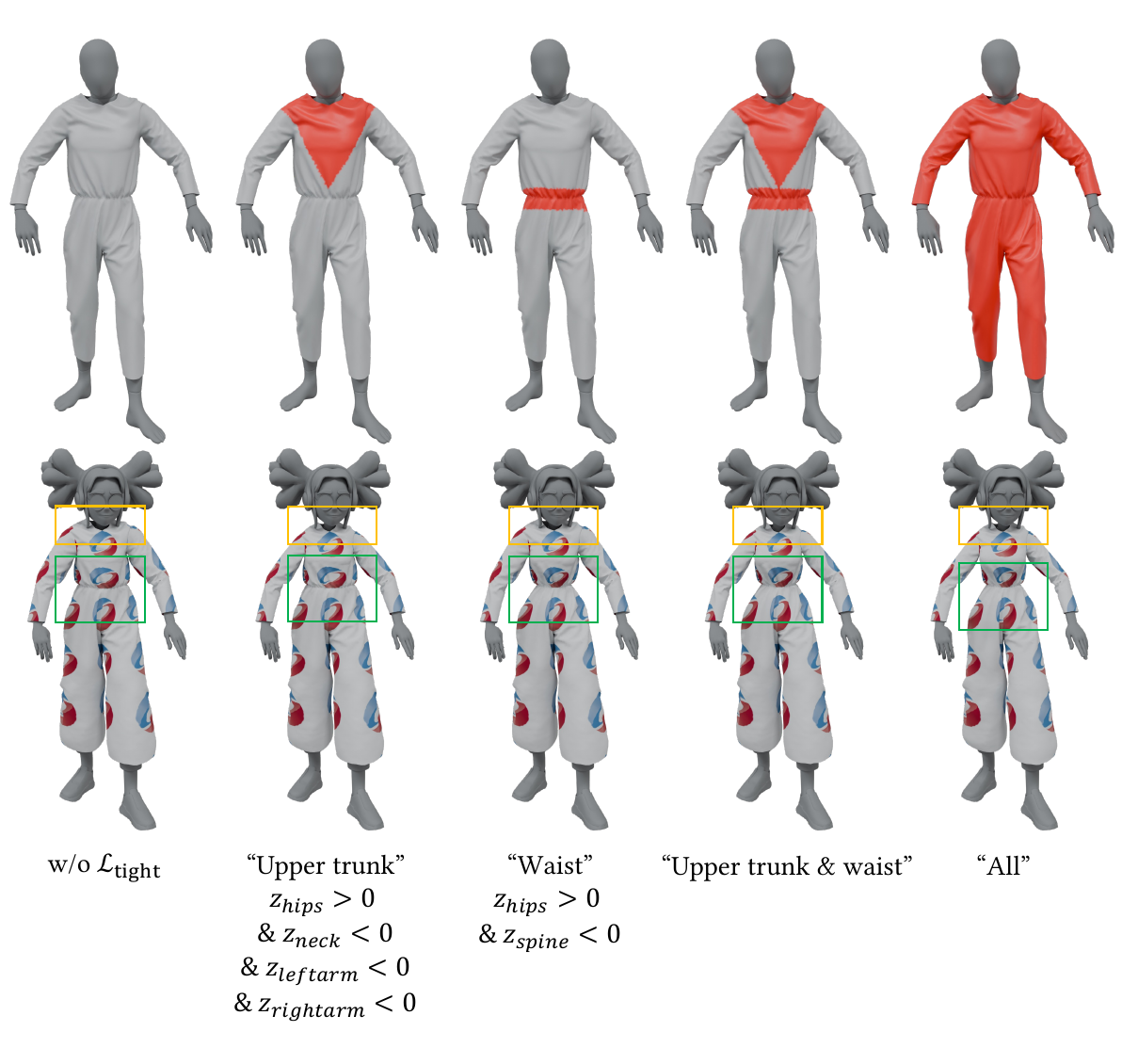}
  \vskip -0.1in
  \caption{\textbf{Fit-control Region}.
  %Fit-style control via the fit-control region $\mathcal{F}$. We define $\mathcal{F}$ using the $z$-coordinates of garment vertices in local bone frames, typically for the \emph{hips} and \emph{spine} bones, to capture tightness around the upper torso and waist. 
  We show the effects of different fit-control settings on the same bodysuit garment. In the first row, we show the same source garment with different user-defined fit regions (highlighted in red). The second row shows LoBoFit refitting results on the same target avatar using the corresponding fit regions from the first row.
  }
  \label{fig:fit-control}
\end{figure}

\section{Implementation Details}
This section provides additional implementation details and practical considerations for reproducing our pipeline. Section \ref{supp: fit-control} describes how we automatically construct fit-control regions from our \emph{LoBoMap} representation to enable intuitive fit-style control. Section~\ref{supp: contact ini} introduces a robust contact-pair initialization strategy that stabilizes optimization under challenging body shapes and articulated poses. Section~\ref{supp: c2f} details our coarse-to-fine optimization schedule for efficient and stable convergence. Finally, Section~\ref{supp: extent_multilayer} presents the operational details of extending \emph{LoBoFit} to multi-layer garment refitting, including handling inter-layer attachments.

\subsection{Fit-control Region $\mathcal{F}$} \label{supp: fit-control}
Based on common practice, tightness around the waist and upper torso effectively characterizes garment–body fit style.
In our implementation, the fit-control region $\mathcal{F}$ is primarily determined by the $z$-coordinates of garment vertices in the local bone frames of the \emph{hips} and \emph{spine} bones (both in $B$). Figure~\ref{fig:fit-control} visualizes $\mathcal{F}$ defined by our local bone mappings and its effect on the same garment.
In most cases, the “Upper trunk \& waist” setting produces results similar to “All,” which enforces, for every garment vertex, the signed distance to the target avatar to match that of the source garment to the source avatar. However, “All” is more reliable when the target avatar is substantially thinner than the source and the garment must shrink, as in Figure~\ref{fig:cycle}.
% Based on common practice, the tightness around the waist and upper torso effectively characterizes garment–body fit style. 
% In our implementation, $\mathcal{F}$ is primarily determined by the $z$-coordinates in the local bone frames of the \emph{hips} and \emph{spine} bones (both in ${B}$). Figure~\ref{fig:fit-control} illustrates the fit-control region $\mathcal{F}$ defined by our local bone projections and the corresponding effect on the same garment. Mostly, "Upper trunk \& Waist" effects similar as "All", which enforces the sign distance of all garment vertices with respect to the target avatar same to the source garment wrt. the source avatar. However, "All" is a better option than "Upper trunk \& Waist", when the target avatar is much thinner than the source, that requires shrinking, such as the example in Figure \ref{fig:ablat}.
%We further demonstrate the effect of different fit-control settings on the same garment. 
%

% We define a fit-control region $\mathcal{F}$ from the user’s semantic input, which specifies where the refitted garment should inherit the source fit style (e.g., \emph{upper trunk}, \emph{waist}, or \emph{upper trunk+waist}). Rather than manually annotating per-vertex labels, our local bone projections induce a natural, body-aware partition of the garment into semantic regions. 
\begin{figure}[ht]
  \centering
  %\fbox{\rule{0pt}{4cm}\rule{0.98\columnwidth}{0pt}} % height=4cm, width=0.8\linewidth
  \includegraphics[width=\columnwidth]{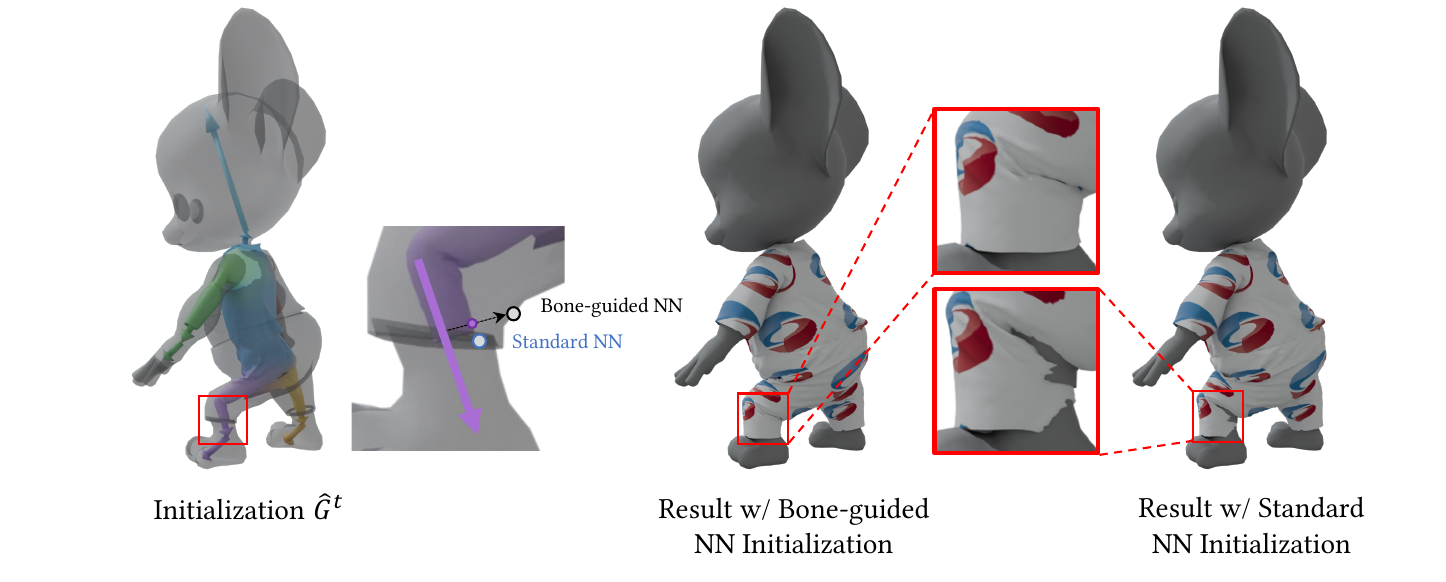}
  \vskip -0.1in
  \caption{\textbf{Effect of bone-guided initialization.}
  Standard nearest-neighbor (Standard NN) search can yield incorrect associations for $\hat{G}^t$, leading to residual garment–body penetrations after refitting. Our bone-guided nearest-neighbor search (Bone-guided NN) produces more reliable initial contacts, resulting in a garment that correctly conforms to the target avatar.
  %For the initialization $\hat{G}^t$, a standard nearest neighbor (Standard NN) query can produce incorrect association. We introduce a bone-guided nearest neighbor (Bone-guided NN) search method to initialize contact pairs, which consequently produces garment refitting results with garment correct conform to the target avatar, whereas standard NN based initialization results some penetration left. 
  }
  \label{fig:contact_ini}
\end{figure}
\subsection{Contact-pair initialization} \label{supp: contact ini}
To compute $\mathcal{L}_{\text{contact}}$, both for penalizing penetrations and enforcing tightness, we construct contact pairs $(g^t, a^t)$, where $g^t$ is a target-garment vertex and $a^t$ is its closest point on the target body $A^t$. For the initialization $\hat{G}^t$, a naive nearest-point query can produce incorrect associations under large shape deviations. While our bone-local 
$z$-coordinate regularization often corrects such errors, it can fail on short-limbed avatars where the $z$ variation is insufficient to disambiguate nearby parts (e.g., the mouse leg in Figure \ref{fig:contact_ini}).
%especially in articulated limbs: on short-limbed avatars (e.g., the mouse), vertices near a bent knee may snap to the opposite leg when the legs are close \Zhang{(Fig.~XX)}.

To improve robustness, we use a bone-guided initialization. For each vertex $\hat g^t$, we find its most related bone $b^*$ as the bone segment with the smallest distance to $\hat g^t$, cast a ray through $\hat g^t$ orthogonal to $b^*$, intersect it with $A^t$, and take the closest valid hit as the initial contact $a^t$ with respect to $\hat g^t$.
We use this strategy \emph{only for initialization}; due to the stability and robustness of our method, the garment reaches the correct surface neighborhood within a few iterations, after which we switch to standard nearest-point updates of the contact pairs, {implemented via an approximate nearest-face search based on KNN over target-body face centroids}. Figure~\ref{fig:contact_ini} illustrates the procedure and its impact.
%We apply this \emph{only at initialization}; after a few iterations the garment moves near the correct surface, and we thereafter update contact pairs using the standard nearest-point query.

%\Zhang{After this initialization stage, once the garment moves sufficiently close to the target body, contact pairs are updated every 2000 iterations using an approximate nearest-face search based on KNN over target-body face centroids. This update strategy provides a good balance between accuracy and computational efficiency.}

\subsection{Coarse-to-Fine Optimization} \label{supp: c2f}
\begin{figure}[ht]
  \centering
  %\fbox{\rule{0pt}{4cm}\rule{0.98\columnwidth}{0pt}} % height=4cm, width=0.8\linewidth
  \includegraphics[width=\columnwidth]{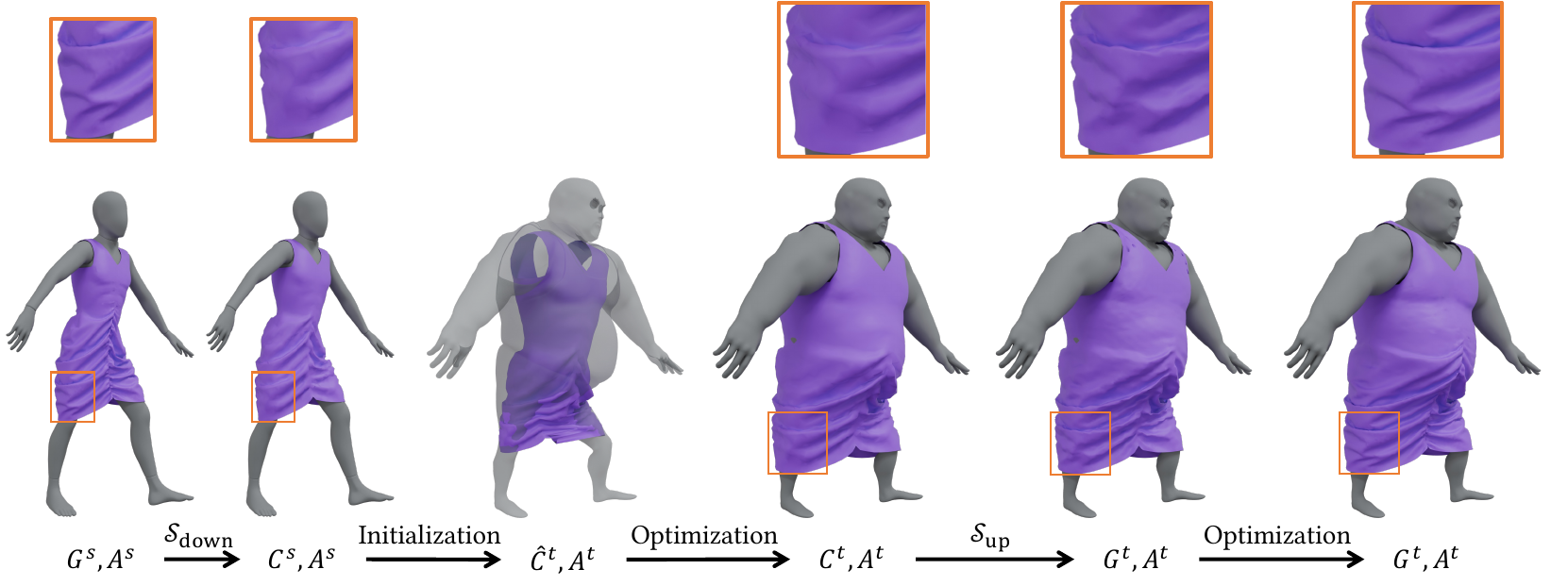}
  \vskip -0.1in
  \caption{\textbf{Coarse-to-fine optimization procedure}.
  We first downsample the source garment $G^s$ to a coarse proxy $C^s$ via $\mathcal{S}_{\text{down}}$ and run our LoBoFit to initialize $\hat{C}^t$ and optimize it into a fitted coarse garment $C^t$ that conforms to the target avatar $A^t$. We than upsample $C^t$ with $\mathcal{S}_{\text{up}}$ to get the high-resolution garment $G^t$ and and further refine it to recover fine-scale details resembling those of the high-resolution source garment.
  }
  \label{fig:coarse-to-fine}
\end{figure}
For high-resolution garments with complex geometry and fine wrinkles, directly optimizing at the target resolution often yields good results, but can be less reliable in challenging cases—occasionally getting trapped in suboptimal local minima and producing artifacts—and is also computationally expensive. We therefore adopt a hierarchical coarse-to-fine optimization procedure, which improves both robustness and efficiency.
We illustrate this procedure in Figure \ref{fig:coarse-to-fine}). 

Following \cite{zhang2024neural}, given a source garment $G^s$, we first leverage the alignment in 2D UV space to establish correspondences between low- and high-resolution garment meshes. We define sampling operators $\mathcal{S}_{\text{down}}$ and $\mathcal{S}_{\text{up}}$, which respectively generate a low-resolution mesh $C$ by downsampling a high-resolution garment $G$, and a high-resolution mesh $G$ by upsampling a low-resolution garment $C$.

We begin by downsampling the source garment $G^s$ to obtain a low-resolution garment $C^s$ using $\mathcal{S}_{\text{down}}$. Following the garment refitting algorithm described in {Section 5}, we then compute a low-resolution refitting initialization $\hat C^t$.
%and the initial garment--body contact pairs $\{\hat c^t, a^t\}$ via bone-related garment--body attachment detection. 
We iteratively deform $C^t$ to obtain a low-resolution garment that fits the target avatar $A^t$, using the objective $\mathcal{L}$ defined with respect to the downsampled source garment $C^s$.
Next, we apply $\mathcal{S}_{\text{up}}$ to upsample $C^t$ and obtain a high-resolution target garment $G^t$. We then further refine $G^t$ by iterative optimization using the same objective $\mathcal{L}$, now defined with the high-resolution source garment $G^s$. Finally, we obtain a target garment $G^t$ that reproduces fine wrinkle details closely resembling those of the high-resolution source garment $G^s$.
Note that we use the AdamW optimizer with a learning rate of $5 \times 10^{-3}$ in the coarse stage. In the high-resolution stage, since the garment already fits the target avatar reasonably well, we reduce the learning rate to $1 \times 10^{-3}$. 
%We showcase the effect of coarse-to-fine optimization in \Zhang{Figure xxx}.

\subsection{Extension to Multi-layer Garments} \label{supp: extent_multilayer}
\begin{figure}[ht]
  \centering
  %\fbox{\rule{0pt}{4cm}\rule{0.98\columnwidth}{0pt}} % height=4cm, width=0.8\linewidth
  \includegraphics[width=\columnwidth]{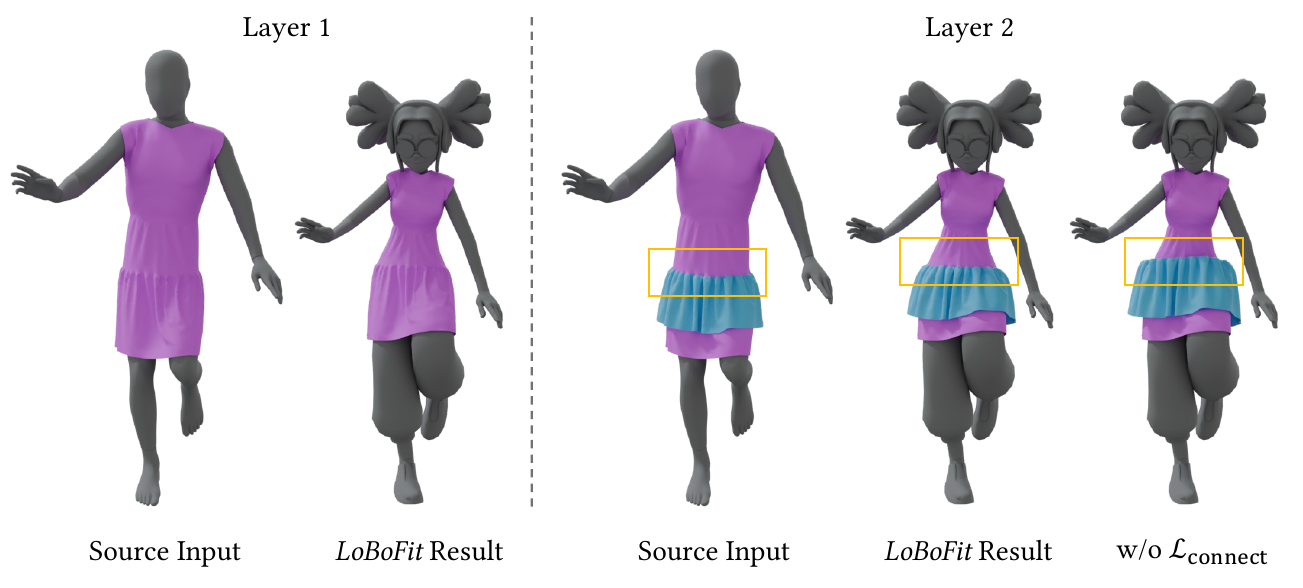}
  \vskip -0.1in
  \caption{\textbf{Garment Refitting multi-layer garment}. 
  We refit layers from inner to outer. For topologically connected layers, we add $\mathcal{L}_{\text{connect}}$ to preserve stitching; without it (w/o $\mathcal{L}_{\text{connect}}$), the outer skirt detaches and appears to float over the inner layer.
  %We refit the layers sequentially from the innermost to the outermost. For layers that are topologically connected, we introduce an additional connection loss term $\mathcal{L}_{\text{connect}}$. We refit garment layers sequentially from the innermost to the outermost. For topologically connected layers, we introduce an additional connection term, $\mathcal{L}_{\text{connect}}$, to preserve inter-layer attachments. Without $\mathcal{L}_{\text{connect}}$, the outer skirt can detach and appear to float over the inner layer instead of remaining stitched.
  }
  \label{fig:multi-layer}
\end{figure}
For multi-layer garments, we refit the layers sequentially from the innermost to the outermost (Figure \ref{fig:multi-layer}). After refitting a high-resolution inner layer, we treat it as part of the target avatar geometry and apply our hierarchical coarse-to-fine optimization to the next layer, thereby also resolving intersections between garment layers. 
For layers that are topologically connected, we introduce an additional connection loss term $\mathcal{L}_{\text{connect}}$ into the contact term $\mathcal{L}_{\text{contact}}$:
\begin{align}
\mathcal{L}_{\text{connect}} &= \sum_{(i,j)} \bigl(\ell^t_{ij} - \ell^s_{ij}\bigr)^2, \\
\mathcal{L}_{\text{contact}} &= \mathcal{L}_{\text{sep}} + \mu_t\,\mathcal{L}_{\text{tight}} + \mu_c\, \mathcal{L}_{\text{connect}},
\end{align}
where $\mu_c$ is a weighting coefficient (set to $\mu_c{=}100$), $(i,j)$ denotes a contact correspondence across layers (vertex $i$ on one layer and $j$ on another), and $\ell^s_{ij}=\|g^s_i-g^s_j\|_2$ is their distance on the source garment $G^s$. We precompute $\ell^s_{ij}$ and use $\mathcal{L}_{\text{connect}}$ to enforce the same inter-layer distances $\ell^t_{ij}$ on the target garment $G^t$.
% where $\mu_c$ is a weighting coefficient (that we set $\mu_c=100 $ in our experiments), $(i,j)$ denotes a pair of corresponding vertices that are in contact across different layers (i.e., $i$ lies on one layer and $j$ on another), and
% $
% \ell^s_{ij} = \| g^s_i - g^s_j \|_2
% $
% is the distance between the layer contact vertex pair $(i,j)$ on source garment mesh $G^s$. We precompute the reference distances $\ell_{ij}^s$ on the source multi-layer garment $G^s$, and enforce the same inter-layer connection distances $\ell_{ij}^t$ on the target garment $G^t$ via $\mathcal{L}_{\text{connect}}$. 
We also demonstrate the effect of $\mathcal{L}_{\text{connect}}$ in Figure \ref{fig:multi-layer}.
%\Zhang{We demonstrate the refitting of multi-layer garments with topological connections between layers to several different target avatars.}

%\section{Experiments}
\begin{figure}
  \centering
  %\fbox{\rule{0pt}{4cm}\rule{\columnwidth}{0pt}} % height=4cm, width=0.8\linewidth
  \includegraphics[width=\columnwidth]{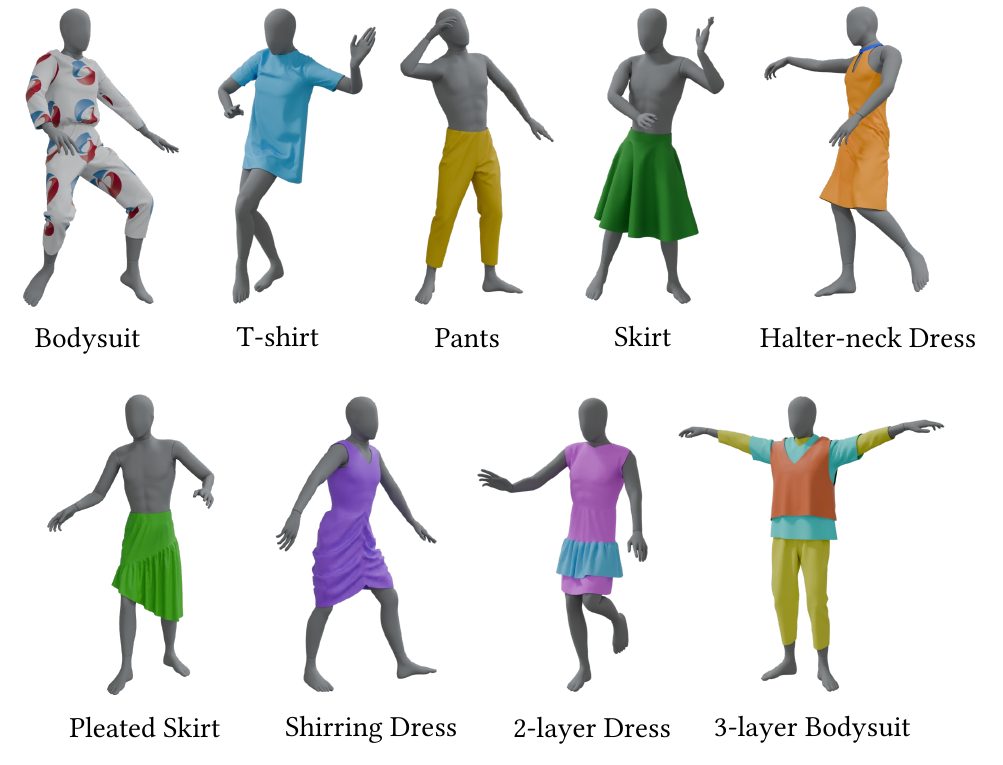}
  \vskip -0.1in
  \caption{\textbf{Data Library}.
  We create nine garments in Marvelous Designer, including a bodysuit, T-shirt, pants, skirt, halter-neck dress, pleated skirt, shirred dress, a two-layer dress, and a three-layer bodysuit; some are adapted from examples in Marvelous Designer’s licensed General Library.
  }
  \label{fig:garm_lib}
\end{figure}

\begin{table}[h]
  \centering
  \caption{\textbf{Data details.}
  We report, for each garment in our experiments, the vertex count, the semantic inputs used to define the fit-control region $\mathcal{F}$(“U” for upper trunk, “W” for waist), and the runtime (in seconds).
  %We reproduce this table (also in the main paper), reporting the mean per-vertex distance between the original garment $G^s$ and the reconstructed garment $\tilde{G}^s$: $\mathrm{mean}\left(\|g^s-\tilde{g}^s\|_2\right)$. We additionally list the semantic inputs used to construct the fit-control region $\mathcal{F}$, applied consistently across methods.
  }
  \label{tab:data_detail}
  \vskip -0.1in
  \begin{tabular}{|c||c|c|c|}
  \hline
   & \# of vertices & $\mathcal{F}$ & Time cost\\
   \hline
   Bodysuit & 27,928 & "U \& W" & 54.11 \\
   \hline
   T-shirt & 16,058 & "U" & 53.49 \\
   \hline
   Pants & 14,630 & "W" & 48.69 \\
   \hline
   Skirt & 20,198 & "W" & 46.81 \\
   \hline
   Halter-neck Dress & 15,038 & "U \& W" & 56.75 \\
   \hline
   Pleated Skirt & 14,225 & "W" & 45.52 \\
   \hline
   Shirring Dress & 19,808 & "U \& W" & 54.41 \\
   \hline
   2-layer Dress & 15,809 & "U \& W" & 57.12 \\
                & +6,819 & $\mathcal{L}_{\text{connect}}$ & +40.68 \\
   \hline
   3-layer Bodysuit & 28,293 & "U \& W" & 63.22 \\
                    & +15,290 & "U" & +48.69 \\
                    & +8,710 & "U" & +47.26 \\
   \hline
  \end{tabular}
  %\vspace{0.5em}
  %\fbox{\parbox{0.9\columnwidth}{\centering \vspace{1cm} Table goes here \vspace{1cm}}}
\end{table}

\section{Data Details}
To evaluate the effectiveness of \emph{LoBoFit}, we download five rigged avatars (Mannequin, Mousey, Jennifer, Michelle, and Ortiz) from Mixamo and additionally sample one male and one female body shape from SMPL~\cite{SMPL:2015}. We treat Mannequin as the source avatar and design all source garments on it, while the remaining avatars serve as targets. All avatars are rigged in Mixamo and animated using two motion sequences (\emph{Wave Hip Hop Dance} and \emph{Catwalk}). We sample poses from \emph{Wave Hip Hop Dance}, excluding those that exhibit severe self-intersections in either the source or target avatars, and use \emph{Catwalk} to evaluate dynamic garment refitting. We extract 20 bones and the corresponding skinning weights in Blender; to support local bone-frame propagation, we further add four auxiliary bones (left/right rib and left/right crotch) to complete the parent–child hierarchy. We create nine garments in Marvelous Designer: a bodysuit, T-shirt, pants, skirt, halter-neck dress, pleated skirt, shirrring dress, a 2-layer dress, and a 3-layer bodysuit. Some of these are adapted from garment examples in the General Library provided by a licensed version of Marvelous Designer. Figure~\ref{fig:garm_lib} summarizes our garment library, with additional details reported in {Table~\ref{tab:data_detail}}.

We also report the runtime of LoBoFit for refitting different garments from the source avatar (Mannequin) to the same target avatar (Ortiz) in Table~\ref{tab:data_detail}, together with the semantic inputs used to define the fit-control region $\mathcal{F}$.
All results use the same coarse-to-fine schedule and fixed optimization settings (6,000 iterations on the coarse proxy and 4,000 iterations on the full-resolution mesh).
We observe modest runtime variations across garments, which is correlated with $\mathcal{F}$: larger fit-control regions involve more vertices in the tightness term $\mathcal{L}_{tight}$, increasing per-iteration cost. 
For example, using “upper trunk” or “upper trunk \& waist” typically takes longer than using “waist” alone.
Overall, for single-layer garments with 10k–30k vertices, LoBoFit consistently runs in roughly one minute.

\section{Baseline Comparisons}
\begin{wrapfigure}{R}{0.25\columnwidth}
\centering
\includegraphics[width=0.25\columnwidth]{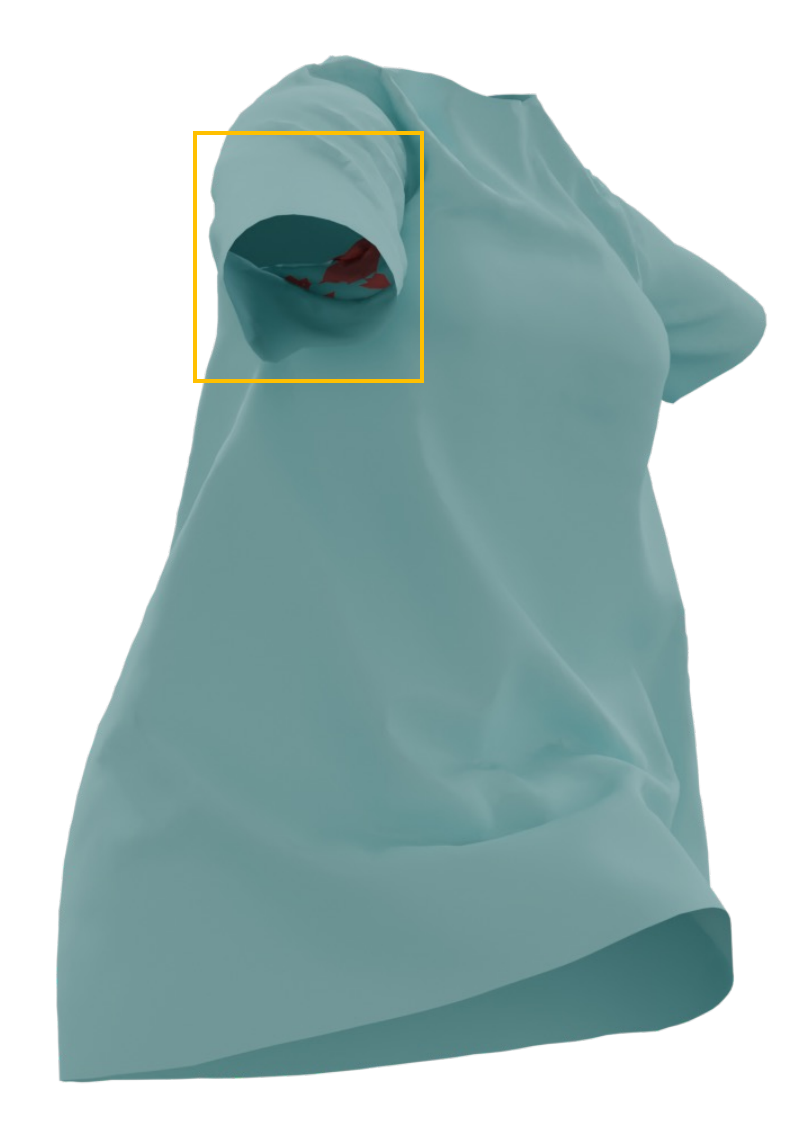}
\end{wrapfigure}
\textbf{Self-intersection analysis.}
{Table~\ref{tab:self_intersection} reports the quantitative comparison of self-intersection with IFGR ~\cite{huang2025intersection}. As expected, the IPC-based method of IFGR handles self-collision more effectively and achieves self-intersection-free results in all tested cases. By contrast, our method is not explicitly designed for self-collision handling and does not include a dedicated self-collision loss.
Nevertheless, our method still produces favorable results in practice. Since \emph{LoBoFit} preserves source geometry and fine-scale wrinkle structures through local detail-aware regularization, it can also maintain a similarly clean configuration when the source garment is free of self-intersection. For example, on \emph{pose\_1} with \emph{Ortiz}, our result is also free of self-intersection, matching IFGR. The remaining intersections mainly occur near armpits, where competing objectives, such as intersection avoidance and wrinkle preservation, may locally conflict (see inset figure). Overall, the ratio of self-intersecting triangles remains below 0.5\% in all cases included in this evaluation, indicating that our method still achieves strong geometric quality even without explicit self-collision handling.}

\begin{table}[h]
\caption{
\textbf{Self-intersection comparison.} We evaluate on the T-shirt mesh with 31,890 triangles. 
All values are reported as the percentage of triangles involved in self-intersections (lower is better).
Owing to its IPC-based formulation, IFGR~\cite{huang2025intersection} achieves self-intersection-free results (0.00\%) in all cases. Our method, \emph{LoBoFit}, does not explicitly model self-collision, yet keeps the self-intersection ratio below 0.5\% across all cases included in this evaluation.
}
\label{tab:self_intersection}
\centering
\begin{tabular}{l l l l}
\hline
{Pose} & {Avatar} & {\emph{LoBoFit}} & {IFGR} \\
\hline
\multirow{3}{*}{\emph{pose\_1}} 
& Ortiz    & 0.00\% & 0.00\% \\
& Michelle & 0.02\% & 0.00\% \\
& Mousey    & 0.14\% & 0.00\% \\
\hline
\multirow{3}{*}{\emph{pose\_2}} 
& Ortiz    & 0.39\% & 0.00\% \\
& Michelle & 0.49\%          & 0.00\% \\
& Mousey    & 0.25\% & 0.00\% \\
\hline
\end{tabular}
\end{table}

\begin{figure}[h]
  \centering
  %\fbox{\rule{0pt}{4cm}\rule{0.98\columnwidth}{0pt}} % height=4cm, width=0.8\linewidth
  \includegraphics[width=\columnwidth]{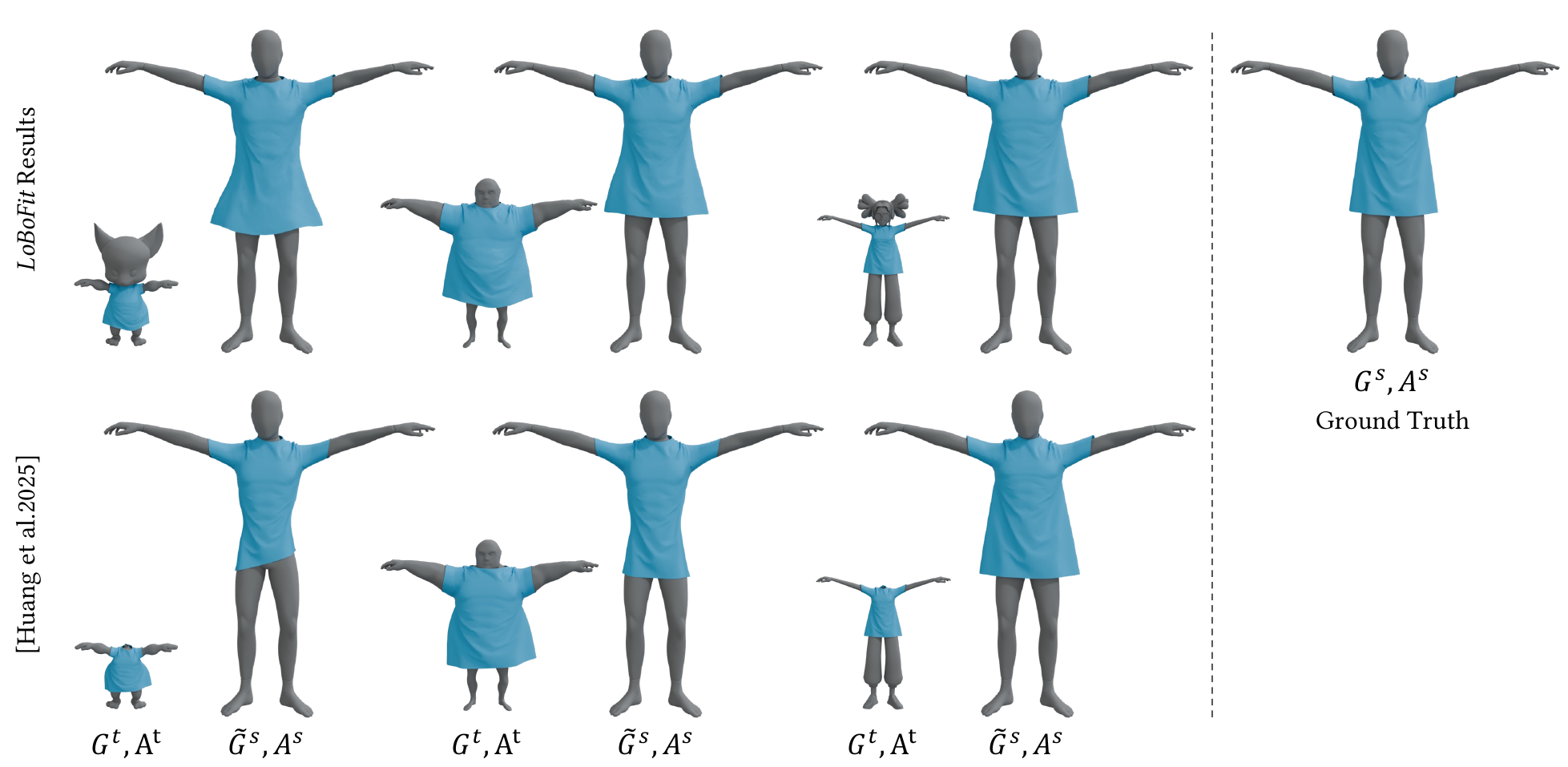}
  \vskip -0.1in
  \caption{\textbf{Cycle consistency evaluation}.
  We evaluate cycle consistency by reversing retargeting, $G^s \rightarrow G^t \rightarrow \tilde{G}^s$. Compared to IFGR, our method better preserves source features, producing $\tilde{G}^s$ that more closely matches the ground-truth $G^s$.
  }
  \label{fig:cycle}
\end{figure}

\textbf{Cycle Consistency.}
We quantitatively compare with IFGR via a cycle-consistency evaluation by reversing the retargeting, $G^s \rightarrow G^t \rightarrow \tilde{G}^s$, where $\tilde{G}^s$ should conform to the source avatar $A^s$. Table~\ref{tab:cyc_eval} summarizes the mean per-vertex error between the original and reconstructed garments, $\mathrm{mean}(\|g^s-\tilde g^s\|_2)$, where \emph{LoBoFit} consistently achieves lower error, indicating better source-feature preservation. Figure~\ref{fig:cycle} visualizes the reverse-retargeting results.
For a fair comparison, we use the same fit-style specification for both methods. In particular, we construct the fit-control region $\mathcal{F}$ using our \emph{LoBoMap} paritioning from user's semantic input, and apply it consistently across methods. In the first two columns of Figure~\ref{fig:cycle}, the (cycle) source avatar is thinner than the target, so we use “All” (i.e., $\mathcal{F}$ includes all garment vertices) to ensure the garment shrinks appropriately.
Recall that fit control in \emph{LoBoFit} is implemented as an explicit signed-distance matching term $|\tilde d^s - d^t|^2$, encouraging vertices in $\mathcal{F}$ to approach a preferred offset $d^t$ derived from the source pair $(G^t, A^t)$. In contrast, IFGR’s IPC-based formulation relies on barrier-type inequality constraints for collision avoidance (e.g., $\tilde d^s \ge 0$), making our fit-style objective nontrivial to integrate without substantial reformulation and re-tuning. In practice, when “All” is used, IFGR often over-tightens the garment, whereas \emph{LoBoFit} maintains the intended fit while better preserving the source design and wrinkle details.

% \begin{table}[h]
% \caption{
% \textbf{Self-intersection comparison.} We evaluate on the T-shirt mesh with 31,890 triangles. 
% All values are reported as the percentage of triangles involved in self-intersections (lower is better).
% Owing to its IPC-based formulation, IFGR~\cite{huang2025intersection} achieves self-intersection-free results (0.00\%) in all cases. Our method, \emph{LoBoFit}, does not explicitly model self-collision, yet keeps the self-intersection ratio below 0.5\% across all cases included in this evaluation.
% }
% \label{tab:self_intersection}
% \centering
% \begin{tabular}{l l l l}
% \hline
% {Pose} & {Avatar} & {\emph{LoBoFit}} & {IFGR} \\
% \hline
% \multirow{3}{*}{\emph{pose\_1}} 
% & Ortiz    & 0.00\% & 0.00\% \\
% & Michelle & 0.02\% & 0.00\% \\
% & Mousey    & 0.14\% & 0.00\% \\
% \hline
% \multirow{3}{*}{\emph{pose\_2}} 
% & Ortiz    & 0.39\% & 0.00\% \\
% & Michelle & 0.49\%          & 0.00\% \\
% & Mousey    & 0.25\% & 0.00\% \\
% \hline
% \end{tabular}
% \end{table}

\begin{table}[h]
  \centering
  \caption{\textbf{Cycle consistency evaluation.}
  We report the mean per-vertex distance between the original garment $G^s$ and the reconstructed garment $\tilde{G}^s$: $\mathrm{mean}\left(\|g^s-\tilde{g}^s\|_2\right)$. We additionally list the semantic inputs used to construct the fit-control region $\mathcal{F}$, applied consistently across methods.
  }
  \label{tab:cyc_eval}
  \vskip -0.1in
  \begin{tabular}{|c||c|c|c|}
  \hline
   & Mousey & Ortiz & Michelle \\
   \hline
   $\mathcal{F}$ & "All" & "All" & "Upper trunk" \\
   \hline
   \emph{LoBoFit} & \textbf{2.41e-2} & \textbf{1.28e-2} & \textbf{1.82e-2} \\
   \hline
   IFGR & 3.61e-2 & 2.36e-2 & 2.76e-2 \\
  \hline
  \end{tabular}
  
  %\vspace{0.5em}
  %\fbox{\parbox{0.9\columnwidth}{\centering \vspace{1cm} Table goes here \vspace{1cm}}}
\end{table}

%Additional qualitative comparisons are provided in Figure~X.

\begin{figure}[ht]
  \centering
  %\fbox{\rule{0pt}{4cm}\rule{0.98\columnwidth}{0pt}} % height=4cm, width=0.8\linewidth
  \includegraphics[width=\columnwidth]{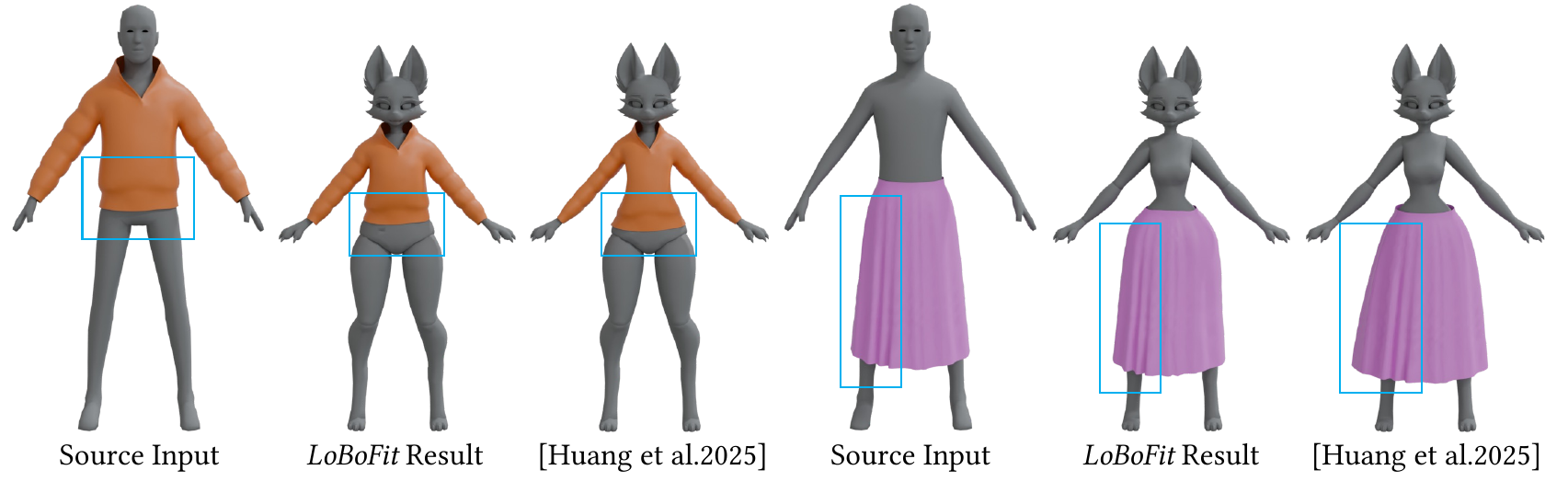}
  \vskip -0.1in
  \caption{\textbf{Baseline Comparisons}.
  We provide additional comparisons with IFGR~\cite{huang2025intersection}, evaluating both IFGR and \emph{LoBoFit} on the same source inputs and target avatars from IFGR. Compared to IFGR, \emph{LoBoFit} more faithfully preserves the source design features in the retargeted garments.
  }
  \label{fig:more}
\end{figure}

\textbf{Additional Comparisons.}
We provide additional comparisons with IFGR in Figure~\ref{fig:more}, evaluating both IFGR and our \emph{LoBoFit} on the source inputs and target avatar provided by IFGR.
{Both methods produce self-intersection-free results in this setting.}
Compared to IFGR, \emph{LoBoFit} produces retargeted garments that more faithfully preserve the source design features.

\clearpage

\end{document}